\documentclass[amsmath,preprintnumbers,aps,nofootinbib,tightenlines,floatfix,superscriptaddress]{revtex4} 

\usepackage{amssymb,latexsym}
\usepackage{amsmath,amsbsy,bbm}
\usepackage{epsfig,bm} 
\usepackage{graphicx,comment}
\usepackage{bm,comment}
\usepackage{epstopdf} \usepackage{color}
\usepackage{subfigure}
\usepackage{slashed,array,mathtools,amsfonts,multirow,hyperref}
\usepackage[utf8]{inputenc}
\DeclareUnicodeCharacter{00A0}{~}

\renewcommand{\v}[1]{\ensuremath{\mathbf{#1}}} 

\newcommand{\avg}[1]{\left< #1 \right>} 

\let\bar=\smallbar 
\newcommand{\bar}[1]{\overline{#1}} 
\let\tilde=\widetilde

\begin{document}

\begin{figure}[!t]
\vskip -1.1cm
\leftline{
\includegraphics[width=3.0 cm]{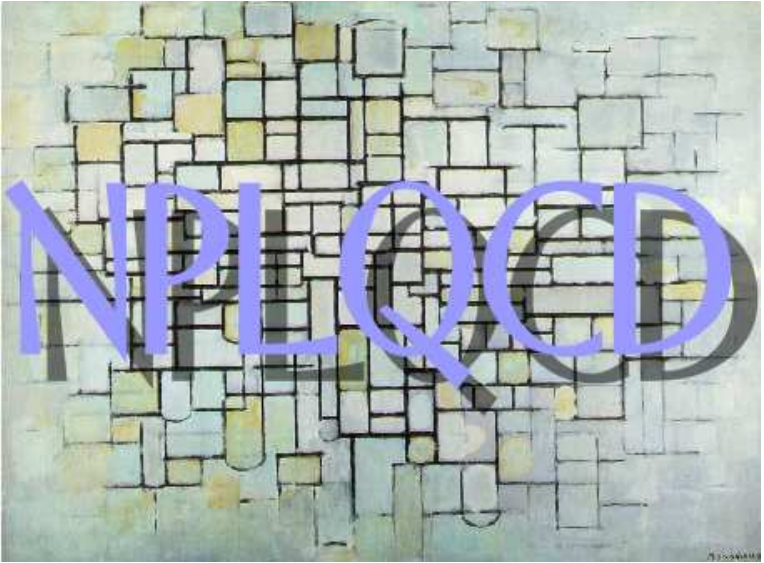}}
\vskip 0.5cm
\end{figure}

\title{On the Statistics of Baryon Correlation Functions in Lattice QCD}

\author{Michael L. Wagman}
\affiliation{%
Institute for Nuclear Theory, Box 351550, Seattle, WA 98195-1550, USA}
\affiliation{%
Department of Physics, University of Washington, Box 351560, Seattle, WA 98195, USA}

\author{Martin J. Savage}
\affiliation{%
Institute for Nuclear Theory, Box 351550, Seattle, WA 98195-1550, USA}


\collaboration{NPLQCD Collaboration}

\date{\today}

\preprint{INT-PUB-16-044}



\begin{abstract}
A systematic analysis of the structure of single-baryon correlation functions calculated with lattice QCD is performed, with a 
particular focus on characterizing the structure of the noise associated with quantum fluctuations.
The signal-to-noise problem in these correlation functions is shown, as long suspected, to result from a sign problem.
The log-magnitude and complex phase are found to be approximately described by normal and wrapped normal distributions respectively.
Properties of circular statistics are used to understand the emergence of a large time noise region where standard energy measurements are unreliable.
Power-law tails in the distribution of baryon correlation functions, associated with stable distributions and ``L{\'e}vy flights'',
are found to play a central role in their time evolution.
A new method of analyzing correlation functions is considered for which the signal-to-noise ratio of energy measurements is constant, rather than exponentially degrading, with increasing source-sink separation time.
  This new method includes an additional systematic uncertainty that can be removed by performing an extrapolation,
  and the signal-to-noise problem re-emerges in the 
  statistics of this extrapolation. 
  It is demonstrated that this new method allows accurate results for the nucleon mass to be extracted from the large-time noise region inaccessible to standard methods.
  The observations presented here are expected to apply to quantum Monte Carlo calculations more generally.
Similar methods to those introduced here may lead to practical improvements in analysis of noisier systems.
 \end{abstract}
\maketitle

\section{Introduction}
\label{sec:introduction}

Modern nuclear physics research relies upon large-scale high-performance computing  (HPC) to predict the 
properties of a diverse array of many-body systems, ranging from the properties of hadrons computed 
from the dynamics of quarks and gluons, 
through to the form of gravitational waves emitted from inspiraling binary neutron star systems.
In many cases,
the entangled quantum nature of these systems and the nonlinear dynamics that
define them, preclude analytic calculation of their properties. 
In these cases, precise numerical evaluations of high-dimensional integrations 
that systematically approach the quantum path integral are required.
Typically, it is average quantities that are determined by Monte Carlo (MC) path integral evaluations.
These average values are 
to be used subsequently in direct comparison with experiment, 
as input to analytic frameworks with outputs that can then be compared with experiment, 
or as predictions for critical components of systems that are inaccessible to experiment 
such as the equation of state of dense matter in explosive astrophysical environments.
Enormous amounts of HPC resources are used in such MC calculations to determine average values of quantities and their uncertainties.
The central limit theorem, and in particular the 
$1/\sqrt{N}$ scaling anticipated for the uncertainties associated with average values, are used to make estimates of projected resource requirements.
When a system has  a ``sign problem'',  for which the average value of a quantity of interest results from cancellations of (relatively) 
large contributions, such as found when averaging $e^{i\theta}$, the HPC resources required for accurate numerical estimates of the average(s) 
are prohibitively large.  
This is the case for numerical evaluations of the path integrals describing strongly interacting systems with even a modest 
non-zero net baryon number.

While the quantum fluctuations (noise) 
of many-body systems contain a wealth of information beyond average values, only a relatively small amount of 
attention has been paid to refining calculations based upon the structure of the noise.
This statement, of course, does not do justice to the fact that all observables (S-matrix elements)
in quantum field theory calculations can be determined from vacuum expectation values of products of quantum fields.
However, in numerical calculations, it is generally the case that  noise is treated as a nuisance,
something to reduce as much as needed, 
as opposed to a feature that may reveal aspects of systems that are obscured through distribution 
among many expectation values.
In the area of Lattice Quantum Chromodynamics (LQCD), 
which is the numerical technique used to evaluate the quantum path integral associated with Quantum Chromodynamics (QCD) that 
defines the dynamics of quarks and gluons, 
limited progress has been made toward understanding the structure of the noise in correlation functions and the physics that it contains.

Strongly interacting quantum systems can be described through path integral representations of correlation functions. In principle, MC evaluation of lattice regularized path integrals can solve QCD as well as many strongly interacting atomic and condensed matter theories. In practice, conceptual obstacles remain and large nuclei and nuclear matter are presently inaccessible to LQCD. 
In the grand canonical formulation, LQCD calculations with non-zero chemical potential face a sign problem where MC 
sampling weights are complex and cannot be interpreted as probabilities. In the canonical formulation, calculations with non-zero baryon number face a Signal-to-Noise (StN) problem where statistical uncertainties in MC results grow exponentially at large times. 
Like the sign problem, the StN problem arises when the sign of a correlation function can fluctuate, at which point cancellations allow for a mean correlation function of much smaller magnitude than a typical MC contribution.

The nucleon provides a relatively simple and well-studied example of a complex correlation function with a StN problem. 
The zero-momentum Euclidean nucleon correlation function $G(t)$ is guaranteed to be real by existence of a Hermitian, bounded transfer matrix and the spectral representation
\begin{equation}
  G(t) = \avg{ C_i(t) } = \sum_{\v{x}} \avg{ N(\v{x},t) \bar{N}(0) } = \sum_{n=0}^\infty \tilde{Z}_n Z^\dagger_n e^{-E_n t} \sim e^{- M_N t}
     \ \ \ ,
  \label{Cdef}
\end{equation}
where $C_i$ denotes an individual nucleon correlation function calculated from quark propagators in the presence of the $i$-th member of a statistical ensemble $U_i$ of $i=1,\dots, N$ gauge field configurations,
$\avg{\cdot}$ denotes an average over gauge field ensembles in $\avg{C_i(t)}$ and an average over quark and gluon fields in the middle term,
$\bar{N}$ and $N$ are nucleon creation and annihilation interpolating operators, 
$\tilde{Z}_n^\dagger$ and $Z_n$ represent the overlap of these interpolating operators onto the $n$-th QCD eigenstates with 
quantum numbers of the nucleon, 
$E_n$ is the energy of the corresponding eigenstate, 
$t$ is Euclidean time, 
$M_N$ is the nucleon mass,
and $\sim$ denotes proportionality in the limit $t\rightarrow \infty$. 
A phase convention for creation and annihilation operators is assumed so that $C_i(0)$ is real for all correlation functions in a statistical ensemble. 
At small times $C_i(t)$ is approximately real, but at large times it must be treated as a complex quantity. 
The equilibrium probability distribution for $C_i(t)$ can be formally defined as
\begin{equation}
  \mathcal{P}\left( C_i(t) \right) = Z^{-1}\  \int \mathcal{D}U\; e^{-S(U)}\delta(C(U;t) - C_i(t))
 \qquad  \ \ {\rm with} \ \  \qquad
  Z\ =\  \int \mathcal{D}U\; e^{-S(U)}
   \ \ \ ,
  \label{correlation functionprobdef}
\end{equation}
where $U$ is a gauge field, $C(U;t)$ is the nucleon correlation function in the presence of a background gauge field 
$U$, $\mathcal{D}U$ is the Haar measure for the gauge group, and $S(U)$ is the gauge action arising after all 
dynamical matter fields have been integrated out. 
For convenient comparison with LQCD results, a lattice regulator 
 with a lattice spacing equal to unity will be assumed throughout. 
Unless specified, results will not depend on details of the ultraviolet regularization of $\mathcal{P}(C_i(t))$.

MC integration of the path integral representation of a partition function, as performed in LQCD calculations, provides a 
statistical ensemble of background quantum fields. 
Calculation of $C_i(t)$ in an ensemble of QCD-vacuum-distributed gauge fields $U_i$ provides a statistical ensemble of correlation functions distributed according to $\mathcal{P}(C_i(t))$. 
Understanding the statistical properties of this ensemble is essential for efficient MC calculations, 
and  significant progress has been achieved in this direction since the early days of lattice field theory. 
Following Parisi~\cite{Parisi:1983ae}, Lepage~\cite{Lepage:1989hd} argued that $C_i(t)$ has a StN problem 
where the noise, or square root of the variance of $C_i(t)$, becomes exponentially larger than the signal, 
or average of $C_i(t)$, at large times. 
It is helpful to review the pertinent details of Parisi-Lepage scaling of the StN ratio.

Higher moments of $C_i(t)$ are themselves field theory correlation functions with well-defined spectral 
representations.~\footnote{
The $n$-th moment $\avg{C_i(t)^n}$ represents the $n$-nucleon nuclear correlation function in the absence of 
Pauli exchange between quarks in different nucleons. 
This is formally a correlation function in a partially-quenched theory with $nN_f$ valence quarks and $N_f$ sea quarks. 
In general, such a theory is guaranteed to have a bounded, but not necessarily Hermitian, transfer matrix~\cite{Bernard:2013kwa}. 
}
Their large-time behavior is a single decaying exponential whose scale is set by the lowest energy state with appropriate quantum numbers. Assuming that matter fields have been integrated out exactly rather than stochastically, $C_i(t)^\dagger C_i(t)$ will contain three valence quarks and three valence antiquarks whose net quark numbers are separately conserved. 
This does not imply that $|C_i(t)|^2$ will only contain nucleon-antinucleon states, as nothing prevents these distinct valence quarks and antiquarks from forming lower energy configurations such as three pions. 
Quadratic moments of the correlation function, therefore, have the asymptotic behavior
\begin{equation}
  \begin{split}
    \avg{C_i(t)^2} \sim e^{-2M_N t}\ \ \ ,
    \hspace{20pt}
     \avg{|C_i(t)|^2} \sim e^{-3m_\pi t}
    \ \ \ \ .
  \end{split}\label{C2Lepage}
\end{equation}
At large times, the nucleon StN ratio is determined by the slowest-decaying moments at large times, taking the form,
\begin{equation}
  \begin{split}
    \frac{\avg{C_i(t)}}{\sqrt{\avg{|C_i(t)|^2}}} \sim e^{-\left( M_N - \frac{3}{2}m_\pi \right)t}
    \ \ \  .
  \end{split}\label{Lepage}
\end{equation}
and is therefore exponentially small.~\footnote{
A generalization of the Weingarten-Witten QCD mass inequalities~\cite{Weingarten:1983uj,Witten:1983ut} 
by Detmold~\cite{Detmold:2014iha} proves that $M_N \geq \frac{3}{2}m_\pi$. 
Assuming that interaction 
energy shifts in the three-pion states contributing to the variance correlation function are negligible,
the nucleon StN ratio is therefore exponentially small for all quark masses. 
}
The quantitative behavior of the variance of baryon correlation function  in 
LQCD calculations was investigated in high-statistics studies by the NPLQCD collaboration~\cite{Beane:2009kya,Beane:2010em,Beane:2014oea} and more recently by Detmold and Endres~\cite{Detmold:2014rfa,Detmold:2014hla}, and was 
found to be roughly consistent with Parisi-Lepage scaling. 
 One of us~\cite{Savage:2010misc} 
 extended Parisi-Lepage scaling to higher moments of $C_i(t)$ and showed that all odd moments 
 of $C_i(t)$ are exponentially suppressed compared to even moments at large times, see Ref.~\cite{Grabowska:2012ik,Beane:2014oea} for further discussion. 
 Nucleon correlation function distributions are increasingly broad and symmetric with exponentially 
 small StN ratios at large times, as seen, for example, in histograms of the real parts of LQCD 
 correlation functions in Ref.~\cite{Beane:2014oea}.

Beyond moments, the general form of correlation function distributions has also been investigated. 
Endres, Kaplan, Lee, and Nicholson~\cite{Endres:2011jm} found that correlation functions in the 
nonrelativistic quantum field theory describing unitary fermions possess approximately log-normal distributions. 
They presented general arguments, that are discussed below, suggesting that this behavior might be a generic feature of 
quantum field theories. 
Knowledge of the approximate form of the correlation function distribution was exploited to construct an improved estimator, the cumulant expansion, that was productively applied to subsequent studies of unitary 
fermions~\cite{Endres:2011er,Endres:2011mm,Lee:2011sm,Endres:2012cw}. 
Correlation function distributions have been studied analytically in the Nambu-Jona-Lasinio 
model~\cite{Grabowska:2012ik,Nicholson:2012xt}, where it was found that real correlation functions were approximately log-normal but complex correlation functions in a physically equivalent formulation of the theory were broad and symmetric at large times with qualitative similarities to the QCD nucleon distribution. 
DeGrand~\cite{DeGrand:2012ik} observed that meson, baryon, and gauge-field correlation functions in 
$SU(N_c)$ gauge theories with several choices of $N_c$ are also approximately log-normal at small times where 
imaginary parts of correlation functions can be neglected. 
These various observations provide strong empirical evidence that the distributions of real correlation functions in 
generic quantum field theories are approximately log-normal.

A generalization of the log-normal distribution for complex random variables that approximately 
describes the QCD nucleon correlation function at large times is presented in this work. 
To study the logarithm of a complex correlation function, it is useful to introduce the magnitude-phase decomposition 
\begin{equation}
  \begin{split}
    C_i(t) = |C_i(t)|e^{i\theta_i(t)} = e^{R_i(t) + i\theta_i(t)}
    \ \ \ \ .
  \end{split}\label{RThdef}
\end{equation}
At small times where the imaginary part of $C_i$ is negligible, previous observations of log-normal correlation 
functions~\cite{DeGrand:2012ik} demonstrate that $R_i$ is approximately normally distributed. It is shown below that 
$R_i$ is approximately normal at all times, and that $\theta_i$ is approximately normal at small times. 
Statistical analysis of $\theta_i$ is complicated by the fact that it is defined modulo $2\pi$. 
In particular, the sample mean of a phase defined on $-\pi < \theta_i \leq \pi$ does not necessarily provide a faithful 
description of the intuitive average phase (consider a symmetric distribution peaked around 
$\pm\pi$ with a sample mean close to zero). 
Suitable statistical tools for analyzing $\theta_i$ are found in the theory of circular statistics and as will 
be seen below that $\theta_i$ is described by an approximately wrapped normal distribution.~\footnote{
See Refs.~\cite{Fisher:1995,Borradaile:2003,Mardia:2009} for 
textbook introductions to circular statistics.
} 
This work is based on a high-statistics analysis of 500,000 nucleon correlation functions 
generated on a single ensemble of gauge-field configurations by the NPLQCD collaboration~\cite{Orginos:2015aya} with LQCD.
This ensemble has a pion mass of $m_\pi \sim 450\text{ MeV}$, physical strange quark mass, 
lattice spacing $\sim 0.12$ fm, and spacetime volume $32^3\times 96$. 
The L{\"u}scher-Weisz gauge action~\cite{Luscher:1984xn} and $N_f = 2+1$ clover-improved 
Wilson quark actions~\cite{Sheikholeslami:1985ij} were used to generate these ensembles,
details of which can be found in Ref.~\cite{Orginos:2015aya}.
Exploratory data analysis of this high-statistics ensemble plays a central role below.

Sec.~\ref{sec:standard} discusses standard statistical analysis methods in LQCD that introduce concepts used below.
In Section~\ref{sec:decomposition}, 
the magnitude-phase decomposition of the nucleon correlation function and connections to the StN problem are discussed. 
Section~\ref{sec:magnitude} describes the distributions of the log-magnitude and its time derivative in more detail,
while Section~\ref{sec:phase} describes the distribution of the complex phase and its time derivative and explains how their features lead to systematic bias in standard estimators during a large-time  region that is dominated by noise. 
Section~\ref{sec:estimator} draws on these observations to propose an estimator for the nucleon mass in which accurate results can be extracted from
  the large-time noise region with a precision that is constant in source-sink separation time $t$ 
but exponentially degrading in an independent time parameter $\Delta t$.
  Section~\ref{sec:conclusion} conjectures about applications to the spectra of generic complex correlation functions and concludes.

\section{Relevant Aspects of Standard Analysis Methods of Correlation Functions}\label{sec:standard}

Typically, in calculations of meson and baryon masses and  their interactions, 
 correlation functions are generated  from combinations of quark- and gluon-level  
sources and sinks with the appropriate hadron-level quantum numbers.  
Linear combinations of these correlation functions are formed, 
either using Variational-Method type techniques~\cite{Luscher:1990ck}, 
the Matrix-Prony technique~\cite{Beane:2009kya},
or other less automated methods,
in order to optimize overlap onto the lowest lying states in the spectrum 
and establish extended plateaus in 
relevant effective mass plots (EMPs).
In the limit of an infinite number of independent measurements, 
the expectation value of the correlation function is a real number at all times, and 
the imaginary part can be discarded as it is known to average to zero. 
The large-time behavior of such correlation functions 
becomes a single exponential (for an infinite time-direction) with an argument determined by the ground-state energy associated with the particular quantum numbers, or more generally the energy of the lowest-lying state with non-negligible overlap.

The structure of the source and sink play a crucial role in determining the utility of sets of correlation functions.
For many observables of interest, it is desirable to optimize the overlap onto the ground state of the system, and to minimize the overlap onto the correlation function dictating the variance of the ground state.
In the case of the single nucleon, the sources and sinks, ${\cal O}$,  
are tuned in an effort to have maximal overlap onto the ground-state nucleon, while minimizing overlap 
of ${\cal O}{\cal O}^\dagger$ onto the three-pion ground state~\cite{Detmold:2014rfa}.
NPLQCD uses  momentum projected hadronic blocks~\cite{Beane:2006mx} generated from quark propagators 
originating from localized smeared sources to suppress the overlap into the three-pion ground state by a factor of 
$1/\sqrt{V}$ where $V$ is the lattice volume, e.g. Ref.~\cite{Beane:2009kya}.  
For such constructions, the variance of the average scales as
$\sim  e^{- 3 m_\pi t}/(V N)$ at large times, 
where $N$ is the number of statistically independent correlation functions,
while the nucleon correlation function scales as $\sim e^{- M_N t}$.
For this set up, the StN ratio scales as $\sim \sqrt{V N} e^{ - (M_N - 3 m_\pi/2) t}$, from which it is clear that 
exponentially large numbers of correlation functions or volumes are required to overcome the StN problem at large times.
The situation is quite different at small and intermediate times in which the variance correlation function is dominated, not by the three-pion ground state, but by the ``connected'' nucleon-antinucleon excited state, which provides a variance contribution that scales as $\sim  e^{- 2 M_N t}/N$.  

This time interval where the nucleon correlation function is in its ground state and the variance correlation function is in a nucleon-antinucleon excited state has been called the ``golden window''~\cite{Beane:2009kya} (GW).
The variance in the GW is generated, in part, by the distribution of overlaps of the source and sink onto the ground state, that differs at each lattice site due to variations in the gluon fields.
In the work of NPLQCD, correlation functions arising from Gaussian-smeared quark-propagator sources and point-like or Gaussian-smeared sinks
that have been used to form single-baryon hadronic blocks.  
Linear combinations of these blocks are 
combined with coefficients 
(determined using the Matrix-Prony technique of Ref.~\cite{Beane:2009kya} or simply by minimizing the 
$\chi^2$/dof in fitting a constant to an extended plateau region)
that
extend the single-baryon plateau region to earlier times, eliminating the contribution from the first excited state of the baryon
and providing access to smaller time-slices of the correlation functions where StN degradation is less severe.
High-statistics analyses of these optimized correlation functions have shown that GW results are exponentially more precise
and have a StN ratio that degrades exponentially more slowly than larger time results~\cite{Beane:2009kya,Beane:2009gs,Beane:2009py} (for a review, see Ref.~\cite{Beane:2010em}).
In particular StN growth in the GW has been shown to be consistent with an energy scale close to zero,
 as is expected from a variance correlation function dominated by baryon, as opposed to meson, states.
Despite the ongoing successes of GW analyses of few-baryon correlation functions,
the GW shrinks with increasing baryon number~\cite{Beane:2009kya,Beane:2009gs,Beane:2009py}
and calculations of larger nuclei may require different analysis strategies suitable for correlation function without a GW.

\begin{figure}[!ht]
	\includegraphics[width=\columnwidth]{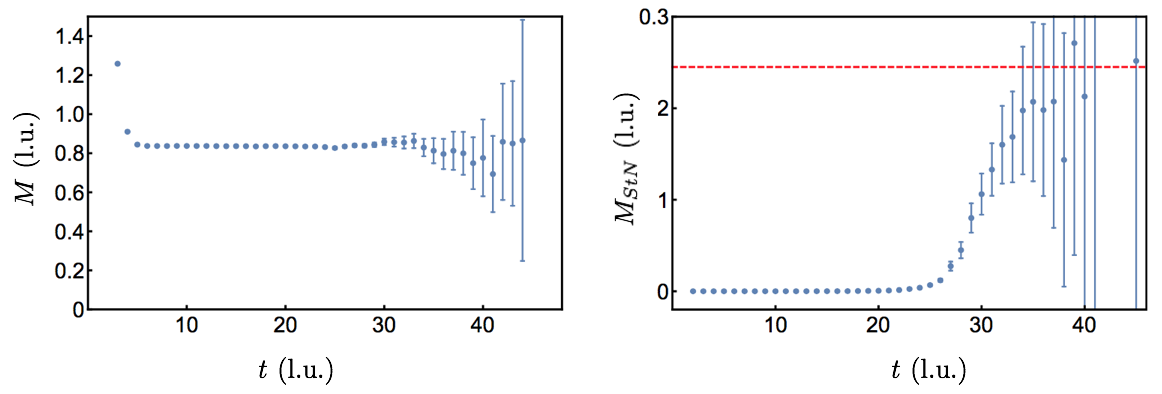}
	\caption{
	\label{fig:xiemp} 
	The EMP associated with the $\Xi$-baryon correlation function with $t_J =2$ (left panel)
	and the energy scale associated with the standard deviation of the ground state energy (right panel).
	This correlation function  is a tuned linear combination 
	of those resulting from localized smeared and 
	point sinks and from a localized smeared source at a pion mass of $m_\pi\sim 450~{\rm MeV}$
	calculated  from 96 sources per configuration on 3538 statistically independent isotropic clover gauge-field 
	configurations~\protect\cite{Orginos:2015aya}.	  
	They have been blocked together to form 100 independent samplings of the combined correlation function.
	The red dashed line in the right panel corresponds to the lowest energy contributing to the StN ratio that is expected to dominate at large times.
	}		
\end{figure}
EMPs, such as that associated with the $\Xi$-baryon shown in Fig.~\ref{fig:xiemp},
 are formed from ratios of correlation functions, which become constant when only 
a single exponential is contributing to the correlation function,
\begin{eqnarray}
  M(t) = \frac{1}{t_J}\ln\left[ {\langle C_i(t) \rangle \over \langle C_i(t+ t_J) \rangle}  \right] & \rightarrow & E_0
\label{eq:emdef}
\ \ \ ,
\end{eqnarray}
where $E_0$ is the ground state energy in the channel with appropriate quantum numbers.
The average over gauge field configurations is typically  over  correlation functions derived from multiple source points on multiple gauge-field configurations.
This is well-known technology and is a ``workhorse'' in the analysis of LQCD calculations.
Typically, $t_J$ corresponds to one temporal lattice spacing, and the jackknife and bootstrap resampling techniques are used to generate covariance matrices in the plateau interval used to extract the 
ground-state energy from a correlated $\chi^2$-minimization~\cite{DeGrand:1990ss,Beane:2010em,Beane:2014oea}.~\footnote{
For pedagogical introductions to LQCD uncertainty quantification with resampling methods, 
see Refs.~\cite{Young,DeGrand:1990,Luscher:2010ae,Beane:2014oea}.
}
The energy can be extracted from an exponential fit to the correlation function or by a direct fit to the effective mass itself.
Because correlation functions generated from the same, and nearby, gauge-field configuration are correlated,
typically they are blocked to form one average correlation function 
per configuration, and  blocked further over multiple configurations, to create an smaller ensemble 
containing (approximately) statistically independent samplings of the correlation function.

\begin{figure}[!ht]
	\includegraphics[width=\columnwidth]{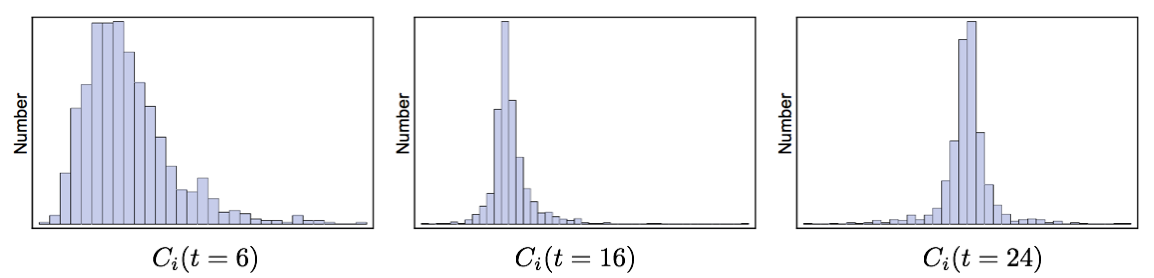}
	\caption{
	\label{fig:ReCdist} 
	The distribution of the real part of
	$10^3$ nucleon correlation functions at time slices $t=6$ (left panel), $t=16$ (middle panel) and $t=24$ (right panel). 
	}		
\end{figure}

It is known that baryon correlation functions contain strong correlations over $\sim m_\pi^{-1}$ time scales, and that these correlations are sensitive the presence of outliers.
Fig.~\ref{fig:ReCdist} shows the distribution of the real part of small-time nucleon correlation functions,
which resembles a heavy-tailed log-normal distribution~\cite{DeGrand:2012ik}.
Log-normal distributions are associated with a larger number of ``outliers'' than arise when sampling a Gaussian distribution,
and the sample mean of these small-time correlation function will be strongly affected by the presence of these outliers.
The distribution of baryon correlation functions at very large source-sink separations is also heavy-tailed;
David Kaplan has analyzed the real parts of NPLQCD baryon correlation functions and found that they resemble a stable distribution~\cite{davidkaplanLuschertalk}.
Cancellations between positive and negative outliers occur in determinations of the sample mean of this large-time distribution,
leading to different statistical issues that are explored in detail in Sec.~\ref{sec:decomposition}.

\begin{figure}[!ht]
  \includegraphics[width=.9\columnwidth]{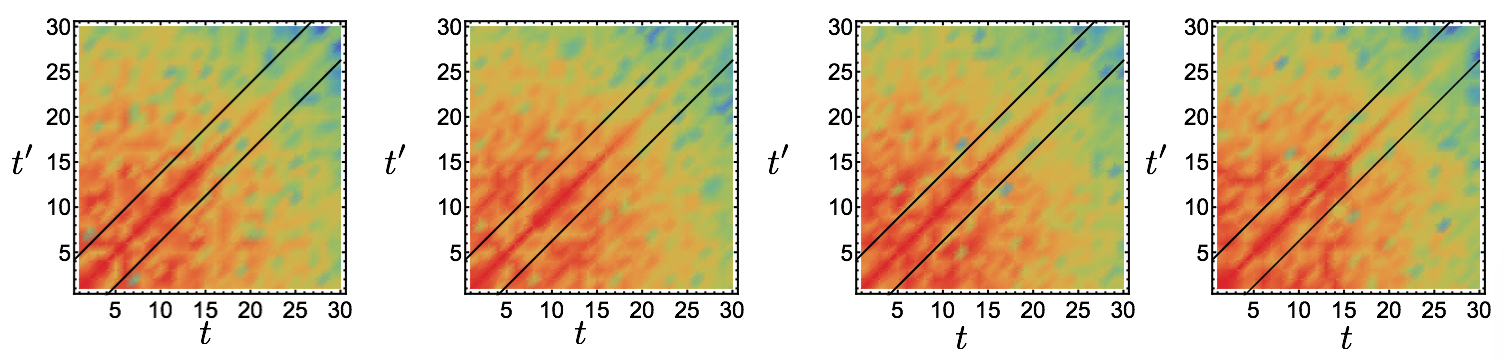}\hspace{5pt}
	\includegraphics[width=.07\columnwidth]{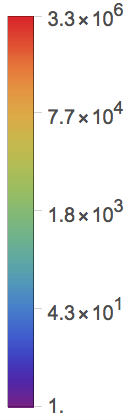}
	\caption{
	\label{fig:jugemean} 
        The logarithm of the inverse covariance matrix determined using booststrap resampling of the sample mean.  
        Lines with $t = m_\pi^{-1}$ and $t^\prime = m_\pi^{-1}$ are shown to demonstrate expected hadronic correlation lengths.
	The correlation function is the same as that described in the caption of Fig.~\protect\ref{fig:xiemp}.
        The normalization of the color scale is identical for all $t_J$.
	}		
\end{figure}

To analyze temporal correlations in baryon correlation functions in more detail,
results for inverse covariance matrices generated through bootstrap resampling of the $\Xi$ baryon effective mass are shown in Fig.~\ref{fig:jugemean}.
The size of off-diagonal elements in the inverse covariance matrix directly sets the size of contributions to the least-squares fit result 
from temporal correlations in the effective mass,
and so it is appropriate to use their magnitude to describe the strength of temporal correlations.
The inverse covariance matrix is seen to possess large off-diagonal elements associated with small time separations
that appear to decrease exponentially with increasing time separation at a rate somewhat faster than $m_\pi^{-1}$.
Mild variation in the inverse covariance matrix is seen when $t_J$ is varied.
Since correlations between $M(t)$ and $M(t^\prime)$ are seen in Fig.~\ref{fig:jugemean} to decrease rapidly as $|t-t^\prime|$ becomes large
compared to hadronic correlation lengths,
is expected that small distance correlations in the covariance matrix decrease
when $C_i(t)$ and $C_i(t-t_J)$ are separated by $t_J \gg m_\pi^{-1}$ and Fig.~\ref{fig:ReCdist},
though such an effect is not clearly visible in the inverse covariance matrix on the logarithmic scale shown.

\begin{figure}[!ht]
  \includegraphics[width=.9\columnwidth]{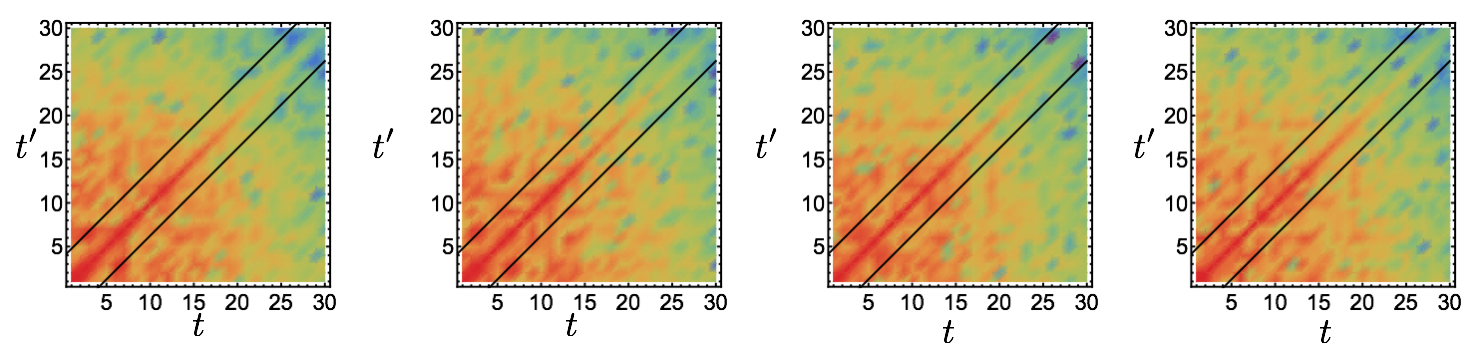} \hspace{5pt}
	\includegraphics[width=.07\columnwidth]{colorscale.png}
	\caption{
	\label{fig:jugeHL} 
	The logarithm of the inverse of the $\Xi$ baryon effective mass covariance matrix for $t_J = 1,2,3,16$ 
	determined using bootstrap resampling of the Hodges-Lehman estimator. 
        Lines with $t = m_\pi^{-1}$ and $t^\prime = m_\pi^{-1}$ are shown to demonstrate expected hadronic correlation lengths.
        The normalization of the color scale is identical for all $t_J$ and further is identical to the normalization of Fig.~\ref{fig:jugemean}.
	}		
\end{figure}

The role of outliers in temporal correlations on timescales $\lesssim m_\pi^{-1}$ is highlighted in Fig.~\ref{fig:jugeHL},
where inverse covariance matrices determined with the Hodges-Lehmann estimator are shown.
The utility of robust estimators, such as the median and the Hodges-Lehmann estimator, with reduced sensitivity to outliers, 
has been explored in Ref.~\cite{Beane:2014oea}.  
When the median and  average of a function are known to coincide,   
there are advantages to using 
the median or Hodges-Lehmann estimator
 to determine the average of a distribution. 
 The associated uncertainty can be estimated with the ``median absolute deviation'' (MAD), and be related to the 
 standard deviation with a well-known scaling factor.
Off-diagonal elements in the inverse covariance matrix associated with timescales $\lesssim m_\pi^{-1}$ are visibly
smaller on a logarithmic scale when the covariance matrix is determined with the Hodges-Lehmann estimator instead of the sample mean.
This decrease in small-time correlations when a robust estimator is employed strongly suggests that
short-time correlations on scales $\lesssim m_\pi^{-1}$ are associated with outliers.

%

\section{A Magnitude-Phase Decomposition}
\label{sec:decomposition}

In terms of the log-magnitude and  phase, the mean nucleon correlation functions is
\begin{equation}
  \begin{split}
    \avg{C_i(t)} =  \int\mathcal{D}C_i\;\mathcal{P}(C_i(t))\; e^{R_i(t) + i\theta_i(t)}
    \ \ \ .
  \end{split}
  \label{Csignproblem}
\end{equation}
In principle, $e^{R_i(t)}$ could be included in the MC probability distribution used for importance sampling. 
With this approach, $R_i(t)$ would contribute as an additional term in a new effective action. 
The presence of non-zero $\theta_i(t)$ demonstrates that this effective action would have an imaginary part. 
The resulting  weight therefore could not be interpreted as a probability and importance sampling 
could not proceed; importance sampling of $C_i(t)$ faces a sign problem. 
In either the canonical or grand canonical approach, one-baryon correlation functions are described by 
complex correlation functions that cannot be directly importance sampled without a sign problem,
but 
 it is formally permissible to importance sample according to the vacuum probability distribution, calculate the  phase resulting from the imaginary effective action on each background field configuration produced in this way, 
 and average the results on an ensemble of background fields. 
 This approach, known as reweighting, has a long history in grand canonical ensemble calculations 
 but has been generically unsuccessful because statistical averaging is impeded by large fluctuations in the complex phase that grow exponentially with increasing spacetime volume~\cite{Gibbs:1986ut,Splittorff:2006fu,Splittorff:2007ck}. 
 Canonical ensemble nucleon calculations averaging $C_i(t)$ over background fields importance sampled 
 with respect to the vacuum probability distribution are in effect solving the sign problem associated with 
 non-zero $\theta_i(t)$ by reweighting. As emphasized by Ref.~\cite{Grabowska:2012ik}, 
 similar chiral physics is responsible for the exponentially hard StN problem appearing in canonical 
 calculations and exponentially large fluctuations of the complex phase in grand canonical calculations.

Reweighting a pure phase causing a sign problem generically produces a StN problem in theories with a mass gap. 
Suppose $\avg{e^{i\theta_i(t)}}\sim e^{-M_\theta t}$ for some $M_\theta \neq 0$. 
Then because $|e^{i\theta_i(t)}|^2 = 1$ by construction, $\theta_i(t)$ has the StN ratio
\begin{equation}
  \begin{split}
    \frac{\avg{e^{i\theta_i(t)}}}{\sqrt{\avg{|e^{i\theta_i(t)}|^2}}} = \avg{e^{i\theta_i(t)}} \sim e^{-M_\theta t}
    \ \ \ ,
  \end{split}\label{ThStN}
\end{equation}
which is necessarily exponentially small at large times. 
Non-zero $M_\theta$ guarantees that statistical sampling of $e^{i\theta_i(t)}$ has a StN problem. 
Strictly, this argument applies to a pure phase but not to a generic complex observable such as 
$C_i(t)$ which might receive zero effective mass contribution from $\theta_i(t)$ and could have important 
correlations between $R_i(t)$ and $\theta_i(t)$. 
MC LQCD studies are needed to understand whether the pure phase StN problem of Eq.~\eqref{ThStN} 
captures some or all of the nucleon StN problem of Eq.~\eqref{Lepage}.

To determine the large-time behavior of correlation functions, it is useful to consider the effective-mass estimator 
commonly used in LQCD spectroscopy, a special case of eq.~(\ref{eq:emdef}), 
\begin{equation}
  \begin{split}
    M(t) = \ln \left[ \frac{\avg{C_i(t)}}{\avg{C_i(t+1)}} \right]
    \ \ \  .
  \end{split}\label{EMdef}
\end{equation}
As $t\rightarrow \infty$, the average correlation function can be described by a single exponential whose  
decay rate is set by the ground state energy, and therefore $M(t)\rightarrow M_N$. 
The uncertainties associated with  $M(t)$ can be estimated by resampling methods such as 
bootstrap. 
The variance of $M(t)$ is generically smaller than that of $\ln\avg{C_i(t)}$ due to cancellations arising from 
correlations between 
$\ln\left[\avg{C_i(t)}\right]$ 
and 
$\ln\left[\avg{C_i(t+1)}\right]$ 
across bootstrap ensembles. 
Assuming that these correlations do not affect the asymptotic scaling of the variance of $M(t)$, 
propagation of uncertainties for bootstrap estimates of the variance of 
$\ln\left[\avg{C_i(t)}\right]$ 
shows that the variance of $M(t)$ scales as
\begin{equation}
  \begin{split}
    \text{Var}\left( M(t) \right) \sim \frac{\text{Var}\left( C_i(t) \right)}{\avg{C_i(t)}^2} \sim e^{2\left( M_N - \frac{3}{2}m_\pi \right)t}
    \ \ \ .
  \end{split}
  \label{EMLepage}
\end{equation}
An analogous effective-mass estimator for the large-time exponential decay of the magnitude is
\begin{equation}
  \begin{split}
    M_R(t) = \ln \left[ \frac{\avg{e^{R_i(t)}}}{\avg{e^{R_i(t+1)}}} \right]
    \ \ \ ,
  \end{split}
  \label{EMRdef}
\end{equation}
and an effective-mass estimator for the  phase is
\begin{equation}
  \begin{split}
    M_\theta(t) = \ln \left[  \frac{\avg{e^{i\theta_i(t)}}}{\avg{e^{i\theta_i(t+1)}}} \right] 
    = 
    \ln \left[ \frac{\avg{\cos(\theta_i(t))}}{\avg{\cos(\theta_i(t+1))}} \right],
  \end{split}
  \label{EMThdef}
\end{equation}
where the reality of the average correlation function has been used.  

\begin{figure}
  \centering
  \includegraphics[width=\columnwidth]{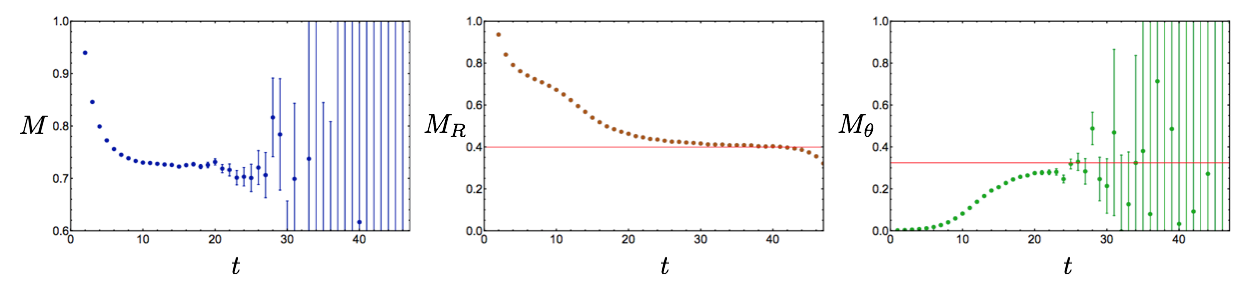}
  \caption{
  The left panel shows the nucleon effective mass $M(t)$ as a function of Euclidean time in lattice units. 
  The middle and right panels show the effective masses $M_R(t)$ and $M_\theta(t)$ of the magnitude and  
  phase respectively. 
  The asymptotic values of $M_R(t)$ and $M_\theta(t)$ are close to $\frac{3}{2}m_\pi$ and $M_N - \frac{3}{2}m_\pi$ respectively, whose values are indicated for comparison with horizontal red lines. 
  The uncertainties are calculated  using bootstrap methods.
  Past $t\gtrsim 30$ the imaginary parts of $\avg{C_i(t)}$ and $\avg{\cos\theta_i(t)}$ are not negligible compared to the real part.
  Here and below we display the real part of the complex log in Eq.~\eqref{EMdef}-\eqref{EMThdef}; taking the real part of the average correlation functions before taking the log or some other prescription would modify the results after $t\gtrsim 30$ in the left and right panels.
  All definitions are equivalent in the infinite statistics limit where $\avg{C_i(t)}$ is real.
  }
  \label{RTh_EM}
\end{figure}
\begin{figure}
  \centering
  \includegraphics[width=\columnwidth]{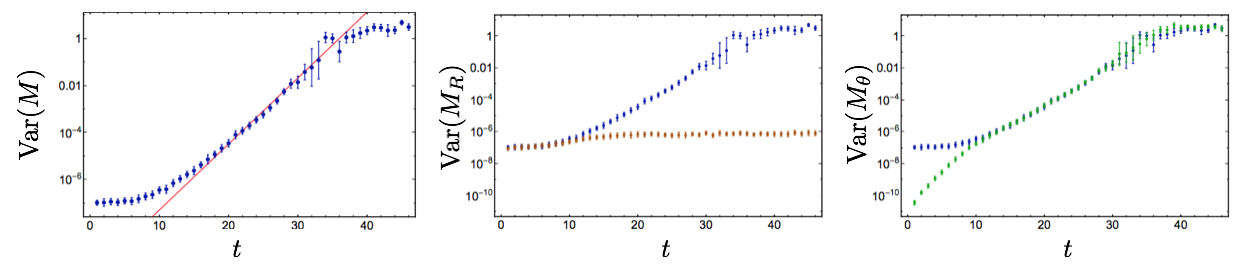}
  \caption{
  Variances of the effective mass estimates shown in Fig.~\ref{RTh_EM}. 
  The blue points common to all panels show the variance of $M(t)$. 
  The red line in the left plot shows a fit to $e^{2(M_N - \frac{3}{2}m_\pi)t}$ variance growth, where the normalization has been fixed to reproduce the observed variance at $t=22$. 
  The orange points in the middle panel show the variance associated with $M_R(t)$.
   The green points in the right panel show the variance associated with $M_\theta(t)$.
   }
  \label{RTh_EMErrors}
\end{figure}

Figure~\ref{RTh_EM} shows EMPs for $M(t)$, $M_R(t)$, and $M_\theta(t)$ calculated from the LQCD ensemble described previously. 
The mass of the nucleon, 
determined from a constant fit in the shaded plateau region $15 \leq t \leq 25$ indicated in Fig.~\ref{RTh_EM},
is found  to be 
$M_N = 0.7253(11)(22)$,
 in agreement with the mass obtained from the golden window in previous studies~\cite{Orginos:2015aya} of 
 $M_N = 0.72546(47)(31)$.
$M_R(t)$ and $M_\theta(t)$ do not visually plateau until much larger times. 
For the magnitude, a constant fit in the shaded region $30 \leq t \leq 40$ gives an effective mass 
$M_R(t) \rightarrow M_R = 0.4085(2)(13)$ which is  close to the value 
$\frac{3}{2}m_\pi = 0.39911(35)(14)$~\cite{Orginos:2015aya} 
indicated by the red line. 
For the  phase, a constant fit to the shaded region $25\leq t \leq 29$ gives an effective mass 
$M_\theta(t) \rightarrow M_\theta = 0.296(20)(12)$, which is  consistent with the value 
$M_N - \frac{3}{2} m_\pi = 0.32636(58)(34)$~\cite{Orginos:2015aya} 
indicated by the red line. 
It is unlikely that the  phase has reached its asymptotic value by this time, but a signal cannot be established at larger times. 
For $t\geq 30$, 
large fluctuations lead to complex effective mass estimates for $M(t)$ and $M_\theta(t)$ and unreliable estimates and uncertainties.  
$M_R(t) + M_\theta(t)$ agrees with $M(t)$ up to $\lesssim 5\%$ corrections at all times, demonstrating that the magnitude and 
cosine of the complex phase are approximately uncorrelated at the few percent level. 
This suggests the asymptotic scaling of the nucleon correlation function can be approximately decomposed as
\begin{equation}
  \begin{split}
    \avg{C_i(t)} \approx \avg{e^{R_i(t)}}\avg{e^{i\theta_i(t)}} \sim \left(  e^{-\frac{3}{2}m_\pi t}\right)\left(  e^{-\left( M_N - \frac{3}{2}m_\pi \right)t}\right)
    \ \ \ \ .
  \end{split}\label{RThScaling}
\end{equation}

For small times $t \lesssim 10$, the means and variances of $M(t)$ and $M_R(t)$ agree up to 
a small contribution from  $M_\theta(t)$. 
This indicates that the real part of the correlation function is nearly equal to its magnitude at small times. 
At intermediate times $10 \lesssim t \lesssim 25$, the contribution of $M_\theta(t)$ grows relative to 
$M_R(t)$, and for $t\gtrsim 15$ the variance of the full effective mass is nearly saturated by the variance of 
$M_\theta(t)$, as shown in Fig.~\ref{RTh_EMErrors}. 
At intermediate times a linear fit normalized to $\text{Var}(M(t=22))$ with slope $e^{2(M_N - \frac{3}{2}m_\pi)t}$ 
provides an excellent fit to bootstrap estimates of $\text{Var}(M(t))$, 
in agreement with the scaling of Eq.~\eqref{EMLepage}. $\text{Var}(M_\theta(t))$ is indistinguishable from 
$\text{Var}(M(t))$ in this region, and $m_\theta(t)$ has an identical StN problem. 
$\text{Var}(M_R(t))$ has 
much more mild time variation, and $M_R(t)$ can be reliably estimated at all times without almost no  StN problem. 
At intermediate times, the presence of non-zero $\theta_i(t)$ signaling a sign problem in importance sampling of $C_i(t)$ appears responsible for the entire nucleon StN problem.

$M(t)$ approaches its asymptotic value much sooner than $M_R(t)$ or $M_\theta(t)$. 
This indicates that the overlap of $\bar{N}(0)N(0)$ onto the three-pion ground state in the variance correlation function is greatly 
suppressed compared to the overlap of $\bar{N}(0)$ onto the one-nucleon signal ground state. 
Optimization of the interpolating operators for high signal overlap contributes to this. 
Another contribution arises from momentum projection, which suppresses the variance overlap factor by 
$\sim 1/(m_\pi^3 V)$~\cite{Beane:2009gs}. 
A large hierarchy between the signal and noise overlap factors provides a GW visible 
at intermediate times $10 \lesssim t \lesssim 25$. 
In the GW, $M(t)$ approaches it's asymptotic value 
but $\text{Var}(M(t))$  begins to grow exponentially and 
$M_\theta(t)$ is suppressed compared to $M_R(t)$. Reliable extractions of $M(t)$ are possible in the GW.

\begin{figure}[!ht]
  \centering
  \includegraphics[width=\columnwidth]{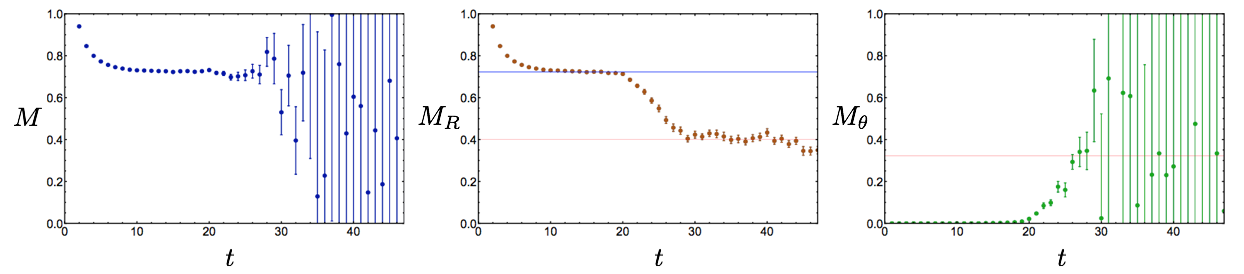}
  \caption{
  EMPs from an ensemble of 500 blocked correlation functions,
  each of which is equal to the sample mean of 1000 nucleon correlation functions. 
  The left panel shows the effective mass $M(t)$ of the blocked correlation functions. 
  The middle panel shows the magnitude contribution $m_R(t)$ and, for reference, 
  a red line at $\frac{3}{2}m_\pi$ and a blue line at $M_N$ are shown. 
  The right panel shows the phase mass $m_\theta(t)$ of the blocked correlation functions 
  along with a red line at $M_N - \frac{3}{2}m_\pi$.
  }
  \label{BlockedRTh_EM}
\end{figure}
\begin{figure}[!ht]
  \centering
  \includegraphics[width=\columnwidth]{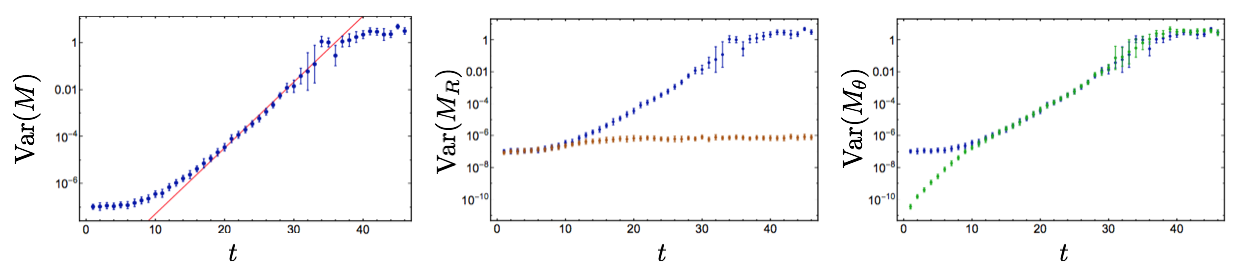}
  \caption{
  Bootstrap estimates of the variance of the effective mass using blocked correlation functions. 
The left panel shows the variance of $M(t)$ for blocked data in blue and the almost indistinguishable variance of $M(t)$ for unblocked data in gray. The middle panel shows the variance of blocked estimates of $m_R(t)$ in orange and the right panel shows the variance of blocked estimates of $m_\theta(t)$ in green.}
  \label{BlockedRTh_EMErrors}
\end{figure}

The effects of blocking, that is averaging subsets of correlation functions and analyzing the distribution of the averages, 
are shown in Fig.~\ref{BlockedRTh_EM}. 
$M_\theta(t)$ is suppressed compared to $M_R(t)$ for larger times in the blocked ensemble, and the log-magnitude 
saturates the average and variance of $M(t)$ through intermediate times $t\lesssim 25$. 
Blocking does not actually reduce the total variance of $M(t)$. 
Variance in $M(t)$ is merely shifted from the phase to the log-magnitude at intermediate times. 
This is reasonable, since the imaginary part of $C_i(t)$ vanishes on average and so blocked correlation functions 
will have smaller imaginary parts. Still, blocking does not affect $\avg{C(t)}$ and only affects bootstrap 
estimates of $\text{Var}(M(t))$ at the level of correlations between correlation functions in the ensemble. 
Blocking also does not delay the onset of a large-time noise region $t\gtrsim 35$ where $M(t)$ and 
$m_\theta(t)$ cannot be reliably estimated.

Eventually the scaling of $\text{Var}(M(t))$ begins to deviate from Eq.~\eqref{EMLepage}, 
and in the noise region $t\gtrsim 35$ the observed variance remains approximately  constant (up to large fluctuations). 
This is inconsistent with Parisi-Lepage scaling. While the onset of the noise region is close to the mid-point of the time direction $t=48$, a qualitatively similar onset occurs at earlier times in smaller statistical ensembles.
Standard statistical estimators therefore do not reproduce the scaling required by basic principles of quantum field theory 
in the noise region. 
This suggests systematic deficiencies leading to unreliable results for standard statistical estimation of correlation 
functions in the noise region. 
The emergence of a noise region where standard statistical tools are unreliable can be understood in terms of the 
circular statistics describing $\theta(t)$ and is explained in Sec.~\ref{sec:phase}. 
A more straightforward analysis of the distribution of $R_i(t)$ is first presented below.

\subsection{The Magnitude}
\label{sec:magnitude}

\begin{figure}[!ht]
  \centering
  \includegraphics[width=\columnwidth]{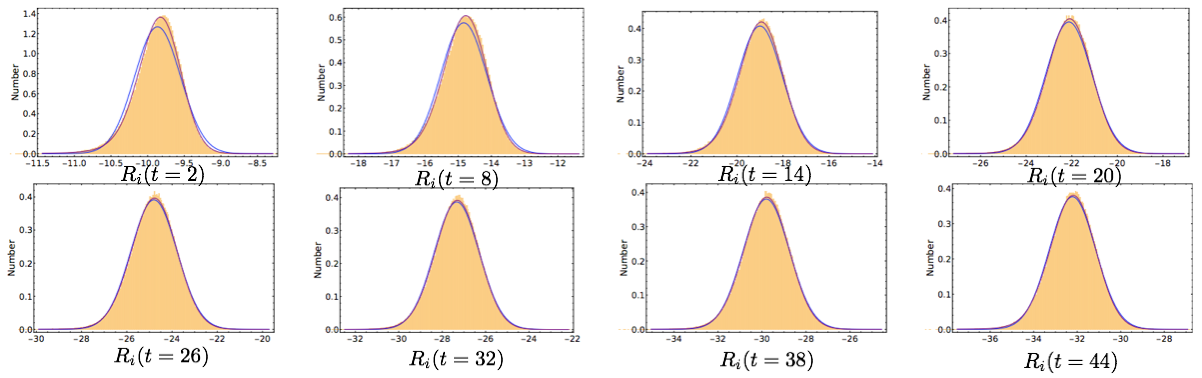}
  \caption{
  Normalized histograms of $R_i(t)$ derived from the LQCD results. 
  The blue curves correspond to  best fit normal distributions  determined from the sample mean and variance,
  while the purple curves correspond to 
  maximum likelihood fits to generic stable distributions. 
  See the main text for more details.
  }
  \label{RHistograms}
\end{figure}

\begin{figure}[!ht]
  \centering
  \includegraphics[width=\columnwidth]{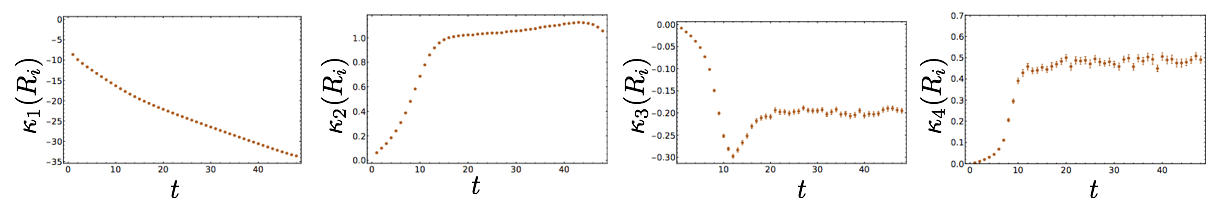}
  \caption{
  The first four cumulants of $R(t)$ as functions of $t$. Cumulants are calculated from sample moments using Eq.~\eqref{cumulantdef} 
  and the associated uncertainties are estimated by bootstrap methods. 
  From left to right, the panels show the cumulants 
   $\kappa_1(R(t))$ (mean),  $\kappa_2(R(t))$  (variance),   $\kappa_3(R(t))$  (characterizing skewness) and $\kappa_4$ (characterizing kurtosis).
  }
  \label{RCumulants}
\end{figure}

Histograms of the nucleon log-magnitude are shown in Fig.~\ref{RHistograms}. 
Particularly at large times, the distribution of $R_i(t)$ is approximately described by a normal distribution. 
Fits to a normal distribution are qualitatively good but not exact, and  deviations between normal distribution fits 
and $R_i(t)$ results are  visible in Fig.~\ref{RHistograms}. 
Cumulants of $R_i(t)$ can be used to quantify these deviations, which can be  recursively calculated  from its moments by
\begin{equation}
  \begin{split}
    \kappa_n\left( R_i(t) \right) = \avg{R_i(t)^n} - \sum_{m=1}^{n-1} {{n-1}\choose{m-1}} \kappa_m\left( R_i(t) \right)\avg{R_i(t)^{n-m}}
    \ \ \  .
  \end{split}\label{cumulantdef}
\end{equation}
The first four cumulants of a probability distribution characterize its mean, variance, skewness, and kurtosis respectively. 
If $|C_i(t)|$ were exactly log-normal, the first and second cumulants of $R_i(t)$, its mean and variance, would fully describe the distribution. 
Third and higher cumulants of $R_i(t)$ would all vanish for exactly log-normal $|C_i(t)|$. 
Fig.~\ref{RCumulants} shows the first four cumulants of $R_i(t)$. 
Estimates of higher cumulants of $R_i(t)$ become successively noisier.

The cumulant expansion of Ref.~\cite{Endres:2011jm} relates the effective mass of a correlation function to the cumulants 
of the logarithm of the correlation function. The derivation of Ref.~\cite{Endres:2011jm} is directly applicable to $M_R(t)$. 
The characteristic function $\Phi_{R(t)}(k)$, defined as the Fourier transform of the probability distribution function of $R_i(t)$, 
can be described by a Taylor series for $\ln[\Phi_{R(t)}(k)]$ whose coefficients are precisely the cumulants of $R_i(t)$,
\begin{equation}
  \begin{split}
    \Phi_{R(t)}(k) = \avg{e^{i k R_i(t)}} = \exp\left[\sum_{n=1}^\infty \frac{(ik)^n}{n!}\kappa_n(R_i(t))\right].
  \end{split}
  \label{cumulants}
\end{equation}
The average magnitude of $C_i(t)$ is given in terms of this characteristic function by
\begin{equation}
  \begin{split}
    \avg{e^{R_i(t)}} = \Phi_{R(t)}(-i) = \exp\left[\sum_{n=1}^\infty \frac{\kappa_n(R_i(t))}{n!}\right].
  \end{split}
  \label{cumulantR}
\end{equation}
This allows application of the cumulant expansion in Ref.~\cite{Endres:2011jm} 
to the effective mass in Eq.~(\ref{EMRdef}) to give,
\begin{equation}
  \begin{split}
    M_R(t) = \sum_{n=1}^\infty \frac{1}{n!}\left[ \kappa_n(R_i(t)) - \kappa_n(R_i(t+1)) \right].
  \end{split}
  \label{cumulantEM}
\end{equation}
Since $\kappa_n(R_i(t))$ with $n >2$ vanishes for normally distributed $R_i(t)$, the cumulant expansion provides a rapidly 
convergent series for correlation functions that are close to, but not exactly, log-normally distributed. 
Note that the right-hand-side of Eq.~\eqref{cumulantEM} is simply a discrete approximation suitable for a lattice regularized 
theory of the time derivative of the cumulants.

\begin{figure}[!ht]
  \centering
  \includegraphics[width=\columnwidth]{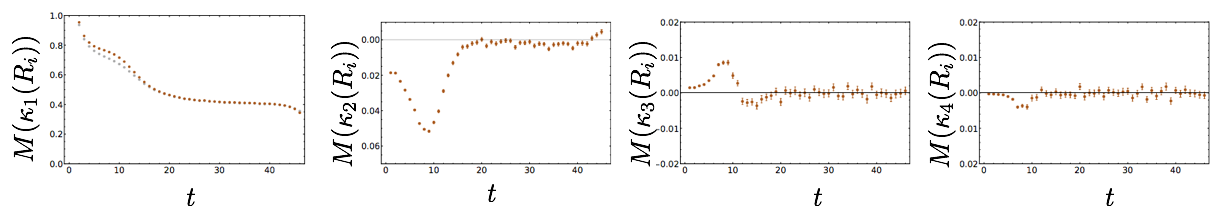}
  \caption{
  Contributions to $M_R(t)$ from the first four terms in the cumulant expansion of Ref.~\cite{Endres:2011jm} 
  given in Eq.~\eqref{cumulantEM}. In the leftmost panel, the gray points correspond to the unapproximated 
  estimate for $M_R(t)$ (that are also shown in  Fig.~\ref{RTh_EM}),
  while the orange points show the contribution from the mean $\kappa_1(R(t))$. 
  The other panels show the contributions to Eq.~\eqref{cumulantEM} associated with the 
  higher cumulants
  $\kappa_2(R_i(t))$, $\kappa_3(R(t))$, and $\kappa_4(R(t))$, respectively.
  }
  \label{RCumulantEM}
\end{figure}
Results for the effective mass contributions of the first few terms in the cumulant expansion of Eq.~\eqref{cumulantEM} 
are shown in Fig.~\ref{RCumulantEM}. 
The contribution $\kappa_1(R_i(t)) - \kappa_1(R_i(t+1))$, 
representing the time derivative of the mean, provides an excellent approximation to $M_R(t)$ after small times. 
$(\kappa_2(R_i(t)) - \kappa_2(R_i(t+1)))/2$ provides a very small negative contribution to $M_R(t)$, and contributions  
from $\kappa_3(R_i(t))$ and $\kappa_4(R_i(t))$ are statistically consistent with zero. 
As $M_R(t)$ approaches its asymptotic value, the log-magnitude distribution can be described to high-accuracy by a 
nearly normal distribution with very slowly increasing variance and small, approximately 
constant  $\kappa_{3,4}$. 
The slow increase of the variance of $R_i(t)$ is consistent with observations above that $|C_i(t)|$ has no severe StN problem.
It is also consistent with expectations that $|C_i(t)|^2$ describes a (partially-quenched) three-pion correlation function with a very mild StN problem, 
with a  scale  set by the attractive isoscalar pion interaction energy.

As Eq.~\eqref{cumulantEM} relates $M_R(t)$ to time derivatives of moments of $R_i(t)$, it is interesting to consider the distribution of the time derivative $\frac{dR_i}{dt}$. 
Defining generic finite differences,
\begin{equation}
  \begin{split}
    \Delta R_i(t, \Delta t) = R_i(t) - R_i(t -  \Delta t)
    \ \ \  ,
  \end{split}
  \label{DeltaRdef}
\end{equation}
the time derivative of lattice regularized results can be defined as the finite difference,
\begin{equation}
  \begin{split}
    \frac{dR_i}{dt} = \Delta R_i(t, 1)
    \ \ \ .
  \end{split}
  \label{dRdtdef}
\end{equation}
If $R_i(t)$ and $R_i(t-1)$ were statistically independent,  it would be straightforward to extract the time derivatives of the 
moments of $R_i(t)$ from the moments of $\frac{dR_i}{dt}$. 
The presence of correlations in time, arising from non-trivial QCD dynamics, obstructs a naive extraction of 
$M_R(t)$ from moments of $\frac{dR)i}{dt}$. 
For instance, without knowledge of $\avg{R_i(t)R_i(t-1)}$ it is impossible to extract the time derivative of the variance of 
$R_i(t)$ from the variance of $\frac{dR_i}{dt}$. 
While the time derivative of the mean of $R_i(t)$ is simply the mean of $\frac{dR_i}{dt}$, 
 time derivatives of the higher cumulants of $R_i(t)$ cannot be extracted from the cumulants of $\frac{dR_i}{dt}$ 
without knowledge of dynamical correlations.

\begin{figure}[!ht]
  \centering
  \includegraphics[width=\columnwidth]{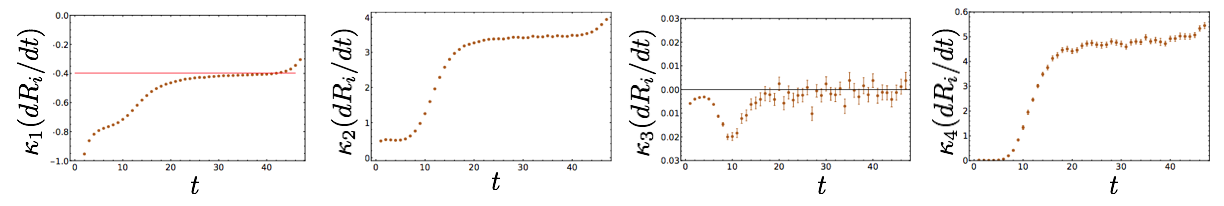}
  \caption{
    The first four cumulants of $\frac{dR_i}{dt}$, determined analogously to the cumulants in Fig.~\ref{RCumulants}.
  }
  \label{dRdtCumulants}
\end{figure}
\begin{figure}[!ht]
  \centering
  \includegraphics[width=\columnwidth]{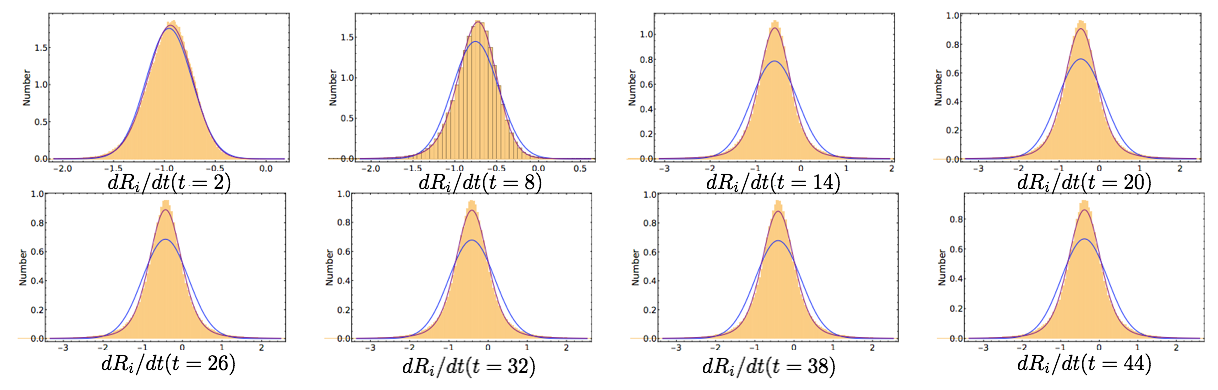}
  \caption{
  Histograms of $\frac{dR}{dt}$, defined as the finite difference $\Delta R(t, 1)$ given in Eq.~(\ref{DeltaRdef}). 
  The blue curves in each panel correspond to the best-fit normal distribution, while the purple curves correspond to the best-fit 
  stable distribution.
  }
  \label{dRdtHistograms}
\end{figure}

The cumulants of $\frac{dR_i}{dt}$ are shown in Fig.~\ref{dRdtCumulants}.  
As expected, the mean of $\frac{dR_i}{dt}$  approaches $\frac{3}{2}m_\pi$ at large times. 
The variance of $\frac{dR_i}{dt}$ is tending to a plateau which is approximately one-third of the variance of $R_i(t)$. 
This implies there are  correlations between $R_i(t)$ and $R_i(t-1)$ that are on the same order of the individual variances 
of $R_i(t)$ and $R_i(t-1)$. 
This is not surprising, given that the QCD correlation length is larger than the lattice spacing. 
No statistically significant $\kappa_3$ is seen for $\frac{dR_i}{dt}$ at large times, but a statistically significant positive $\kappa_4$ is found. 
Normal distribution fits to $\frac{dR_i}{dt}$ are found to be poor, as  shown in Fig.~\ref{dRdtHistograms}, as they 
underestimate both the peak probability and the probability of finding ``outliers'' in the tails of the distribution. 
Interestingly, 
Fig.~\ref{dRdtCumulants}, and histograms of $\frac{dR_i}{dt}$ shown in Fig.~\ref{dRdtHistograms},
suggest that the distribution of $\frac{dR_i}{dt}$ becomes approximately time-independent at large times.

Stable distributions are found to provide a much better description of $\frac{dR_i}{dt}$, and are  consistent with the heuristic arguments 
for log-normal correlation functions given in Ref.~\cite{Endres:2011jm}. 
Generic correlation functions can be viewed as products of creation and annihilation operators with many transfer matrix factors 
describing Euclidean time evolution. 
It is difficult to understand the distribution of products of transfer matrices in quantum field theories, 
but following Ref.~\cite{Endres:2011jm} insight can be gained by considering products of random positive numbers.
As a further simplification, one can consider a product of independent, identically distributed positive numbers, each schematically representing a product of many transfer matrices describing time evolution over a period much larger than all temporal correlation lengths.
Application of the central limit theorem to the logarithm of a product of many independent, identically distributed 
random numbers shows that the logarithm of the product tends to become normally distributed as the number of 
factors becomes large. 
The central limit theorem in particular assumes that the random variables in question 
originate from distributions that have a finite variance.
A generalized central limit theorem proves that sums of heavy-tailed random 
variables tend to become distributed according to stable distributions (that 
include the normal distribution as a special case), 
suggesting that
 stable distributions arise naturally in the logs of products of random variables.

Stable distributions are named as such because their shape is stable under averaging of independent copies of a random variable. 
Formally, stable distributions form a manifold of fixed points in a Wilsonian space of probability distributions where averaging 
independent random variables from the distribution plays the role of renormalization group evolution. 
A parameter $\alpha$, called the index of stability, dictates the shape of a stable distribution and remains fixed under averaging transformations. 
All probability distributions with finite variance evolve under averaging towards the normal distribution, 
a special case of the stable distribution with $\alpha = 2$. 
Heavy-tailed distributions with ill-defined variance evolve towards generic stable distributions with $0 < \alpha \leq 2$. 
In particular, stable distributions with $\alpha < 2$ have power-law tails; 
for a stable random variable $X$ the tails decay as $X^{-(\alpha + 1)}$. 
The heavy-tailed Cauchy, Levy, and Holtsmark distributions are special cases of stable distributions with $\alpha = 1,\;1/2,$ and $3/2$ respectively,
that arise in physical applications.~\footnote{
Further details can be found in  textbooks and reviews on stable distributions and their applications in physics. 
See, for instance, Refs.~\cite{Chandrasekhar:1943,Bouchaud:1990,Bardou:2000,Voit:2005,Nolan:2015} and references within.
}

Stable distributions for a real random variable $X$ are defined via Fourier transform,
\begin{equation}
  \begin{split}
    \mathcal{P}_S(X;\alpha,\beta,\mu,\gamma) &= \int \frac{dk}{2\pi}e^{-i k X}\Phi_X(k;\alpha,\beta,\mu,\gamma)
    \ \ \ ,
  \end{split}
  \label{PSdef}
\end{equation}
of their characteristic functions
\begin{equation}
  \begin{split}
    \Phi_X(k;\alpha,\beta,\mu,\gamma) = \exp\left( i \mu k - |\gamma k|^\alpha\left[1 - i\beta\frac{k}{|k|}\tan(\pi\alpha/2)\right]\right)
    \ \ \  ,
  \end{split}
  \label{PhiSdef}
\end{equation}
where $0<\alpha\leq 2$ is the index of stability, 
$-1\leq \beta \leq 1$ determines the skewness of the distribution, 
$\mu$ is the location of peak probability, 
$\gamma$ sets the width.
For $\alpha=1$,  the above parametrization does not hold and $\tan(\pi\alpha/2)$ should be replaced by $-\frac{2}{\pi}\ln|k|$.
For $\alpha > 1$ the mean is $\mu$, and for $\alpha \leq 1$ the mean is ill-defined. 
For $\alpha =2$ the variance is $\sigma^2 = \gamma^2/2$ and Eq.~\eqref{PhiSdef} implies the distribution is independent of $\beta$,
while
for $\alpha < 2$ the variance is ill-defined.

\begin{figure}[!ht]
  \centering
  \includegraphics[width=\columnwidth]{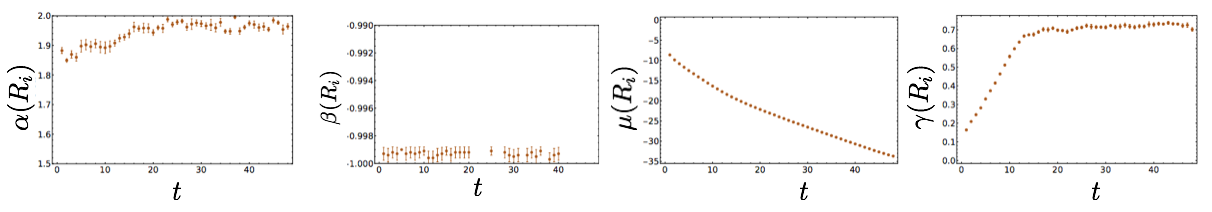}
  \caption{
  Maximum likelihood estimates for stable distribution fits of $R_i(t)$ in terms of the parameters of 
  Eq.~\eqref{PSdef}-\eqref{PhiSdef}. 
  $\alpha=2$ corresponds to a normal distribution. 
  The associated uncertainties are estimated by bootstrap methods.
  Changes in $\beta$ do not affect the likelihood when $\alpha=2$, and reliable estimates of $\beta(R_i(t))$ are not obtained at all times.
  }
  \label{RStable}
\end{figure}
The distributions of $R_i(t)$ obtained from the LQCD calculations
can be fit to stable distributions through maximum likelihood estimation of the stable parameters 
$\alpha,\;\beta,\;\mu,$ and $\gamma$, 
obtaining the results that are shown in Fig.~\ref{RStable}. 
Estimates of $\alpha(R_i)$ are consistent with $2$, corresponding to a normal distribution. 
This is not surprising, because higher moments of $|C_i(t)|$ would be ill-defined and diverge in the infinite statistics 
limit if $R_i(t)$ were literally described by a heavy-tailed distribution. 
$\beta(R_i)$ is strictly ill-defined when $\alpha(R_i)=2$, but results consistent with $\beta(R_i) = -1$ indicate negative 
skewness in agreement with observations above. 
Estimates of $\mu(R_i)$ and $\gamma(R_i)$ are consistent with the cumulant results above if a normal distribution ($\alpha(R_i) = 2$) is assumed. 
Fits of $R(t)$ to generic stable distributions are shown in Fig.~\ref{RHistograms},  and are roughly consistent with  fits to a normal distribution, though some skewness is captured by the stable fits.

\begin{figure}[!ht]
  \centering
  \includegraphics[width=\columnwidth]{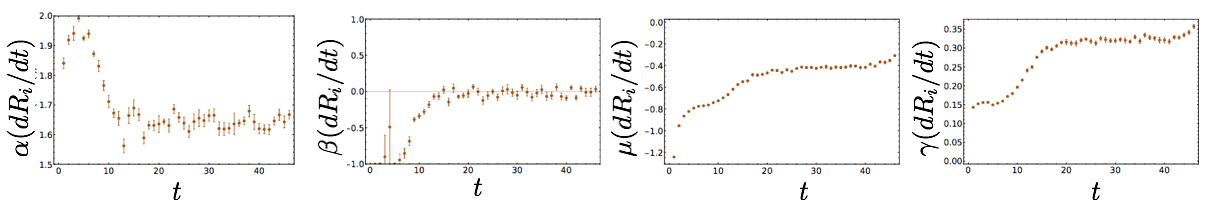}
  \caption{
  Maximum likelihood estimates for stable distribution fits of $\frac{dR_i}{dt}$ similar to Fig.~\ref{RStable}.
  The associated uncertainties are estimated by bootstrap methods.
}
  \label{dRdtStable}
\end{figure}
Stable distribution fits to $\frac{dR_i}{dt}$ indicate statistically significant deviations from a normal distribution ($\alpha=2$), 
as seen in Fig.~\ref{dRdtStable}.
 The large-time distribution of $\frac{dR_i}{dt}$ appears time independent, and fitting $\alpha\left( \frac{dR_i}{dt} \right)$ in the large-time plateau region gives an estimate of the large-time index of stability. Recalling $\frac{dR_i}{dt}$ describes a finite difference over a physical time interval of one lattice spacing, the estimated index of stability is
\begin{equation}
  \begin{split}
    \alpha\left( \Delta R(t\rightarrow \infty, \Delta t \sim 0.12\text{ fm}) \right) \rightarrow 1.639(4)(1).
  \end{split}
  \label{alphaDeltaR}
\end{equation}
Maximum likelihood estimates for $\mu\left( \frac{dR_i}{dt} \right)$ are consistent with the sample mean, and $\beta\left( \frac{dR_i}{dt} \right)$ is consistent with zero in agreement with observations of vanishing skewness.
Therefore,  the distribution of $ \frac{dR_i}{dt}$ is symmetric, as observed in Fig.~\ref{dRdtHistograms}, with power-law tails scaling as 
$\sim \left(\Delta R_i\right)^{-2.65}$ over this time interval of $\Delta t\sim 0.12~{\rm fm}$.

\begin{figure}[!ht]
  \centering
  \includegraphics[width=\columnwidth]{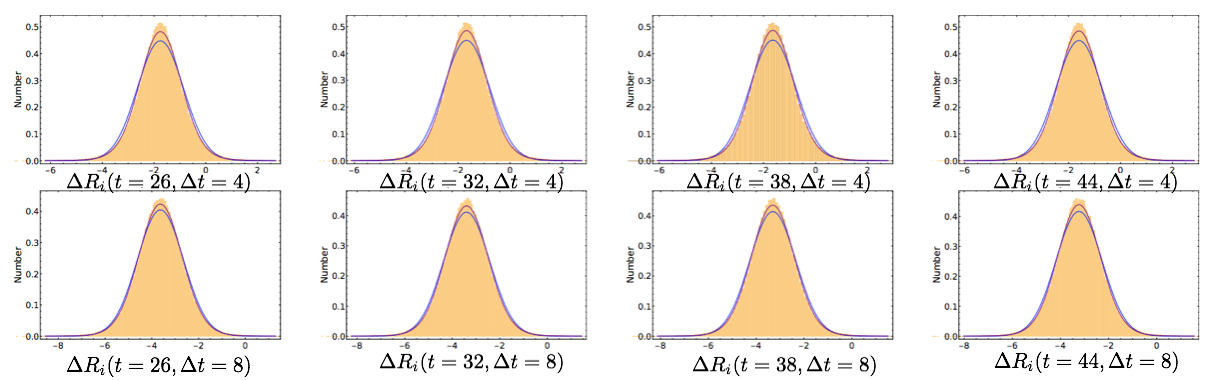}
  \caption{
  Histograms of $\Delta R_i(t, \Delta t)$ for selected large-time values of $t$. 
  The top row shows results for  $\Delta t = 4$, 
  the bottom row shows results for  $\Delta t = 8$,
  and   Fig.~\ref{dRdtHistograms} shows the results for  $\Delta t = 1$. 
  The blue curves represent fits to a normal distribution, while the purple curves represent fits to a stable distribution.
  }
  \label{DeltaRHistograms}
\end{figure}
\begin{figure}[!ht]
  \centering
  \includegraphics[width=\columnwidth]{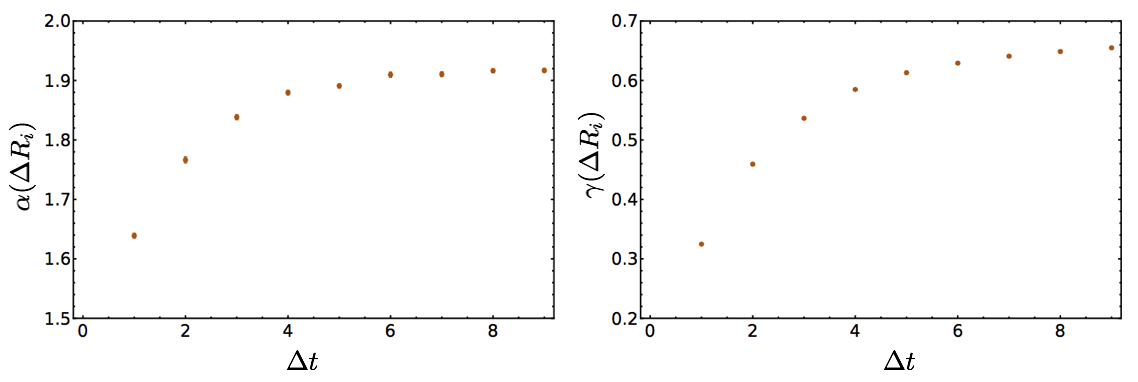}
  \caption{
  Maximum likelihood estimates for the index of stability, $\alpha\left(\Delta R_i(t, \Delta t)\right)$ and width $\gamma\left( \Delta R_i(t, \Delta t) \right)$, in the large-time 
  plateau region as a function of $\Delta t$.  Associated uncertainties are estimated with bootstrap methods.
  }
  \label{DeltaRStable}
\end{figure}

The value of $\alpha\left( \frac{dR_i}{dt} \right)$ depends on the physical time separation used in the finite difference 
definition Eq.~\eqref{DeltaRdef}, and 
stable distribution fits can be performed for generic finite differences $\Delta R_i(t, \Delta t)$. 
For all $\Delta t$, the distribution of $\Delta R_i$ becomes time independent at large times. 
Histograms of the large-time distributions $\Delta R$ for $\Delta t = 4,\;8$ are shown in Fig.~\ref{DeltaRHistograms}, and 
the best fit large-time values for $\alpha\left( \Delta R_i\right)$ and $\gamma\left( \Delta R_i \right)$ are shown in Fig.~\ref{DeltaRStable}. 
Since QCD has a finite correlation length,  $\Delta R_i(t, \Delta t)$ can be described as the difference of 
approximately normally distributed variables at large $\Delta t$. 
In the large $\Delta t$ limit, $\Delta R_i$ is therefore necessarily almost normally distributed,
and correspondingly, $\alpha(\Delta R_i)$, shown in Fig.~\ref{DeltaRStable}, 
increases with  $\Delta t$ and begins to approach the 
normal distribution value $\alpha(\Delta R_i)\rightarrow 2$ for large $\Delta t$. 
A large $\Delta t$ plateau in $\alpha(\Delta R_i)$ is observed that demonstrates small but statistically significant departures from $\alpha(\Delta R_i)< 2$.
This deviation is consistent with the appearance of small but statistically significant measures of non-Gaussianity in $R_i(t)$ seen in Fig.~\ref{RCumulants}.
Heavy-tailed distributions are 
found to be needed only to describe the distribution of $\Delta R_i$ when $\Delta t$ is small enough such 
that $R_i(t)$ and $R_i(t-\Delta t)$ are physically correlated. 
In some sense, the deviations from normally distributed differences, i.e. $\alpha(\Delta R_i) < 2$, are a measure 
the strength of dynamical QCD correlations on the scale $\Delta t$.

\begin{figure}[!ht]
  \centering
  \includegraphics[width=16cm]{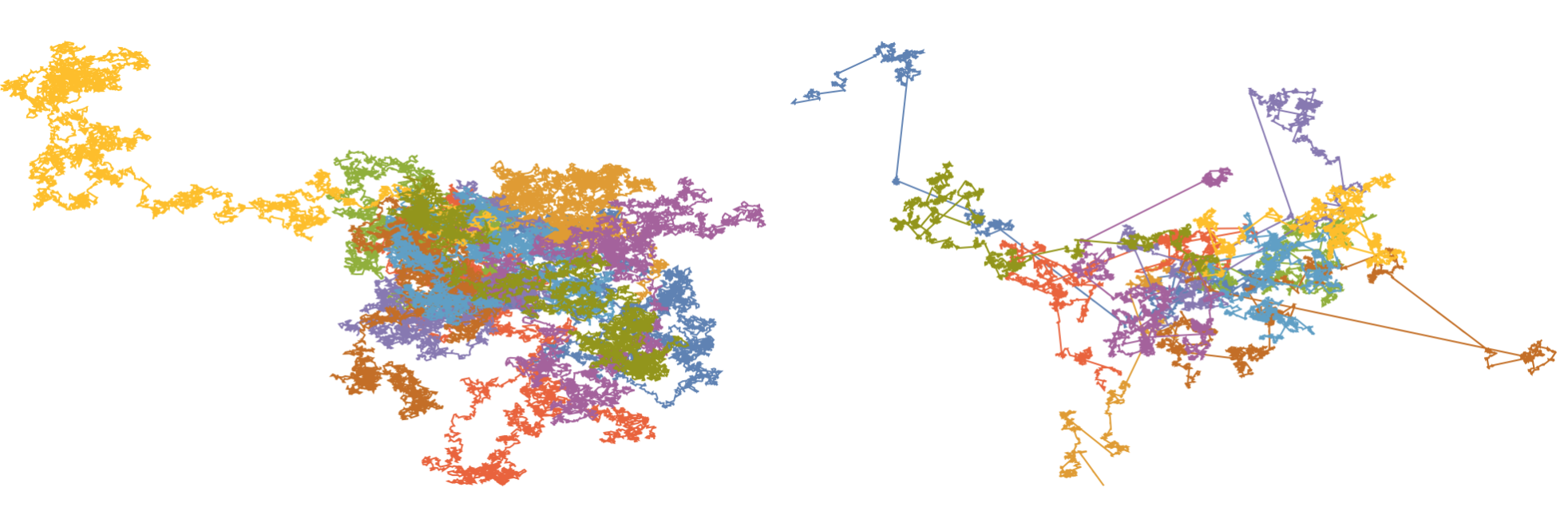}
  \caption{
  The two-dimensional motion of tests particles with their random motion taken from symmetric Stable Distributions.
  At each time step, the angle of the outgoing velocity is chosen randomly with respect to the incident velocity 
  while the magnitude of the velocity is chosen from a symmetric Stable Distribution with $\alpha=2$ 
  corresponding to Brownian motion (left panel), 
  and $\alpha=1.5$ corresponding to a Holtsmark distribution (right panel).
  In the right panel, the large separations between clusters achieved during one time interval correspond to L{\'e}vy flights.
    }
  \label{fig:levyflights}
\end{figure}

The heavy-tailed distributions of $\Delta R_i$ for dynamically correlated time separations correspond to time evolution $\frac{dR_i}{dt}$ 
that is quite  different to that of diffusive Brownian motion  describing the   quantum mechanical motion of free point particles. 
Rather than Brownian motion, heavy-tailed jumps in $R_i(t)$ correspond to a superdiffusive random walk or L{\'e}vy flight. 
Power-law, rather than exponentially suppressed, large jumps give L{\'e}vy flights a qualitatively different character than 
diffusive random walks, including fractal self-similarity, as can be seen in Fig.~\ref{fig:levyflights}.
The dynamical features of QCD that give rise to superdiffusive time evolution are presently unknown, however, 
we conjecture that instantons play a  role.
Instantons are associated with large, localized fluctuations in gauge fields,
and we expect that instantons may also be responsible for infrequent, large fluctuations 
in hadronic correlation functions generating the tails of the $dR_i/dt$ distribution.
It would be interesting to understand if 
 $\alpha\left( \frac{dR_i}{dt} \right)$ can be simply related to observable properties of the nucleon. 
It is also not possible to say from this single study whether $\alpha\left( \frac{dR_i}{dt} \right)$ has a well-defined 
continuum limit for infinitesimal $\Delta t$. 
Further LQCD studies are required to investigate the 
continuum limit of $\alpha\left( \frac{dR_i}{dt} \right)$. 
Lattice field theory studies of other systems and 
calculations of $\alpha\left( \frac{dR_i}{dt} \right)$ in perturbation theory, effective field theory, 
and models of QCD could  provide important insights into the dynamical origin of superdiffusive time evolution.~\footnote{
For example, 
an analysis of  pion correlation functions from the same ensemble of gauge-field configurations
shows that  $R_i$ and ${dR_i\over dt}$ are 
both approximately normally distributed, with $\alpha=1.96(1)$ and 
$\alpha=1.97(1)$,
respectively.
We conclude that the pion shows only small deviations from free particle Brownian motion.
}

One feature of LQCD $\frac{dR_i}{dt}$ results is not well described by a stable distribution. 
The variance of heavy-tailed distributions is ill-defined, and were $\frac{dR_i}{dt}$ truly described by a heavy-tailed 
distribution then the variance and higher cumulants of $\frac{dR_i}{dt}$ would increase without bound as the size of the 
statistical ensemble is increased. 
This behavior is not observed. 
While the distribution of $\frac{dR_i}{dt}$ is well-described by a stable distribution near its peak, the extreme tails of the 
distribution of $\frac{dR_i}{dt}$ decay sufficiently quickly that the variance and higher cumulants of $\frac{dR}{dt}$ 
shown in Fig.~\ref{dRdtCumulants} give statistically consistent results as the statistical ensemble size is varied. 
This suggests that $\frac{dR_i}{dt}$ is better described by a truncated stable distribution, a popular model for,
for example,
financial markets exhibiting high volatility but with a natural cutoff on trading prices, in which some form of sharp cutoff 
is added to the tails of a stable distribution~\cite{Voit:2005}. 
Note that the tails of the $\frac{dR_i}{dt}$ 
distribution describe extremely rapid changes in the correlation function and are sensitive to ultraviolet properties of the theory. 
One possibility is that $\frac{dR_i}{dt}$ describes a stable distribution in the presence of a (perhaps smooth) cutoff arising from ultraviolet regulator effects that damps the stable distribution's power-law decay at very large $\frac{dR_i}{dt}$.
Further studies at different lattice spacings will be 
needed to understand the form of the truncation and whether the truncation scale is indeed set by the lattice scale. 
It is also possible that there is a strong interaction length scale providing a modification to the distribution at large 
$\frac{dR_i}{dt}$,
and it is further possible that stable distributions only provide an approximate description at all $\frac{dR_i}{dt}$.
For now we simply observe that a truncated stable distribution with an unspecified high-scale modification provides a 
good empirical description of $\frac{dR_i}{dt}$.

Before turning to the complex phase of $C_i(t)$, we summarize the main findings about the log-magnitude:
\begin{itemize}
  \item 
  The log-magnitude of the nucleon correlation function in LQCD is approximately normally distributed with small 
  but statistically significant negative skewness and positive kurtosis.
  \item 
  The magnitude effective mass $M_R(t)$ approaches $\frac{3}{2}m_\pi$ at large times, consistent with expectations from Parisi-Lepage scaling for the nucleon variance $|C_i(t)|^2 \sim e^{-3 m_\pi t}$. 
  The plateau of $M(t)$ marks the start of the golden window where excited state systematics are negligible and 
  statistical uncertainties are increasing slowly. 
  The much larger-time plateau of $M_R(t)$ roughly coincides with the plateau of $M_\theta(t)$ to $M_N - \frac{3}{2}m_\pi$ and 
  occurs after variance growth of $M(t)$ reaches the Parisi-Lepage expectation $e^{2(M_N - \frac{3}{2}m_\pi)t}$. 
  Soon after, a noise region begins where the variance of $M(t)$ stops increasing and the effective mass cannot be reliably estimated.
  \item 
  The  log-magnitude does not have a severe StN problem, and $M_R(t)$ can be measured accurately across all 48 timesteps 
  of the present LQCD calculations.  
  The variance of the log-magnitude distribution only increases by a few percent in 20 timesteps after visibly plateauing.
  \item 
  The cumulant expansion describes $M_R(t)$ as a sum of the time derivatives of the cumulants of the log of the correlation function. 
  At large times, the time derivative of the mean of $R_i(t)$ is constant and approximately 
  equal to $M_R(t)$. Contributions to $M_R(t)$ from the variance and higher cumulants of $R_i(t)$ are barely 
  resolved in the sample of $500,000$ correlation functions. 
  \item 
  Finite differences in $R_i(t)$, $\Delta R_i(t, \Delta t)$, are described by time independent distributions at large times. 
  For large $\Delta t$ compared to the QCD correlation length, $\Delta R$ describes a difference of approximately 
  independent normal random variables and is therefore approximately normally distributed. 
  For small $\Delta t$, $\Delta R_i$ describes a difference of dynamically correlated variables. 
  The mean of $\frac{dR_i}{dt}$ is equal to the time derivative of the mean of $R_i(t)$ and therefore provides a good 
  approximation to $M_R(t)$. 
  The time derivatives of higher cumulants of $R_i(t)$ cannot be readily extracted from cumulants of $\frac{dR_i}{dt}$ 
  without knowledge of dynamical correlations.
  \item 
  At large times, $\frac{dR_i}{dt}$ is well described by a symmetric, heavy-tailed, truncated stable distribution. 
  The presence of heavy tails in $\frac{dR_i}{dt}$ indicates that $R_i(t)$ is not described by free particle Brownian motion 
  but rather by a superdiffusive L{\'e}vy flight. 
  Deviations of the index of stability of $\frac{dR_i}{dt}$ from a normal distribution quantify the amount of dynamical correlations 
  present in the nucleon system, the physics of which is yet to be understood. 
  Further studies are required to determine the continuum limit value of the index of stability associated with 
  $\frac{dR_i}{dt}$ and the dynamical origin and generality of superdiffusive L{\'e}vy flights in quantum field theory correlation functions.
\end{itemize}

\subsection{The Phase}
\label{sec:phase}

\begin{figure}[!ht]
  \centering
  \includegraphics[width=\columnwidth]{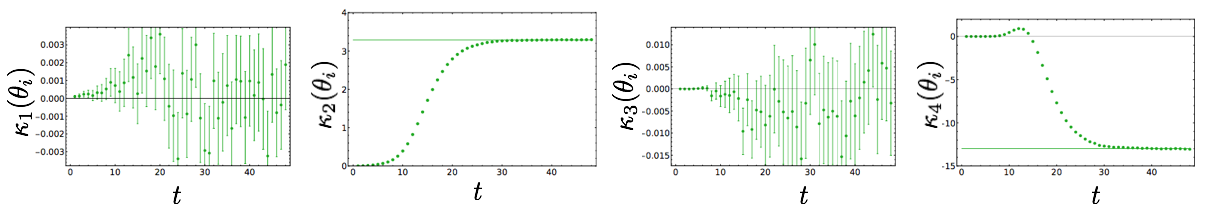}
  \caption{
  The first four cumulants of $\theta_i(t)$. 
  In these fits, no special care is given to the fact that $\theta_i(t)$ is a phase defined on $-\pi<\theta_i(t)\leq \pi$ and standard sample moments 
  are used to determine these cumulants in analogy to Eq.~\eqref{cumulantdef}. 
  Uniform distribution results of $\frac{\pi^2}{3}$ variance and $-\frac{2\pi^4}{15}$ 
  fourth cumulant are shown as green
   lines for reference.
  }
  \label{ThCumulants}
\end{figure}

\begin{figure}[!ht]
  \centering
  \includegraphics[width=\columnwidth]{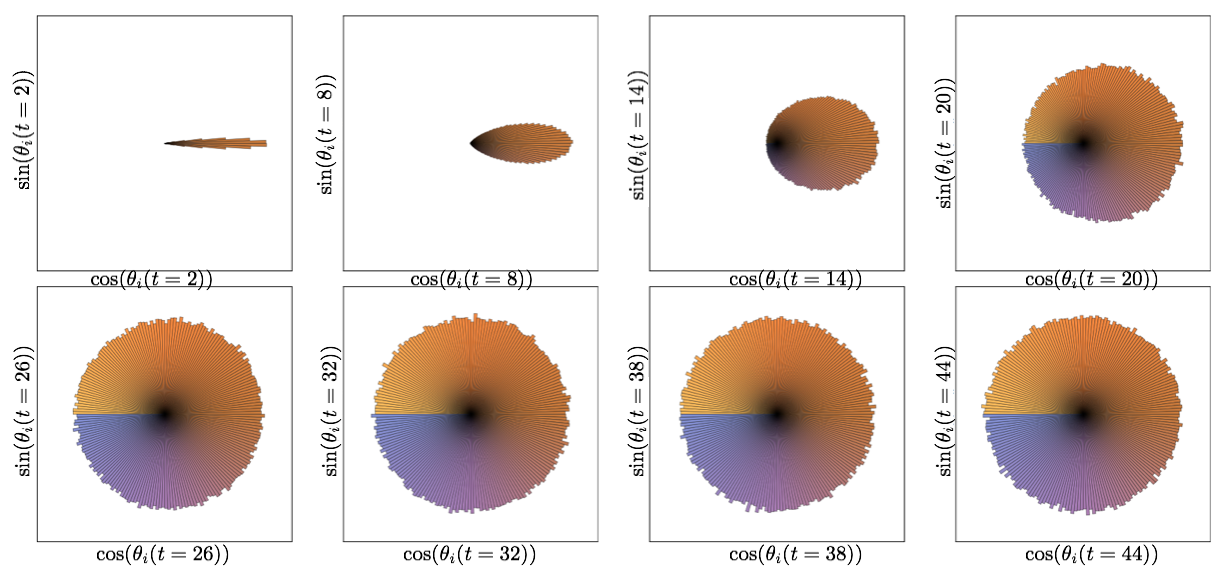}
  \caption{
  Angular histograms of $\theta_i(t)$. 
  The unit circle is split into a uniform sequence of bins, and the number of $\theta_i(t)$ samples falling in 
  each bin sets the radial length of a bar at that angle.
  Colors ranging from orange to blue also denotes angle, and is included to indicate the $\theta_i = \pi$ location of the branch cut in $\theta_i(t) = \text{arg} C_i(t)$.
  }
  \label{ThAngularHistograms}
\end{figure}

The reality of  average correlation functions requires that the distribution of $\theta_i(t)$ be symmetric under 
$\theta_i(t) \rightarrow -\theta_i(t)$. 
Cumulants of $\theta_i(t)$ calculated from sample moments in analogy to Eq.~\eqref{cumulantdef} are shown in Fig.~\ref{ThCumulants}. 
The mean and $\kappa_3$ are noisy but statistically consistent with zero as expected. 
The variance and $\kappa_4$ are small at small times since every sample of $\theta_i(t)$ is defined to vanish at $t=0$, and grow linearly 
at intermediate times $10 < t < 20$ around the golden window. 
After $t=20$, this linear growth slows and they become constant at large times, and are consistent with results from a uniform distribution. 
\begin{figure}[!ht]
  \centering
  \includegraphics[width=\columnwidth]{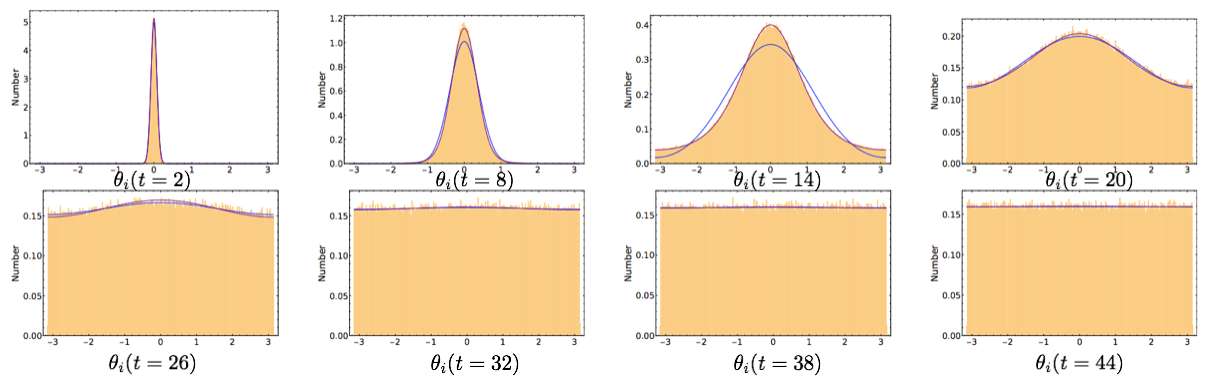}
  \caption{
    Histograms of $\theta_i(t)$ with fits to wrapped normal distributions using Eq.~\eqref{Rdef} shown in blue and fits to wrapped stable distributions using maximum likelihood estimation of the parameters of Eq.~\eqref{PWSdef} shown in purple.
  See the main text for details.
  }
  \label{ThHistograms}
\end{figure}
Histograms of $\theta_i(t)$ shown in Fig.~\ref{ThHistograms} qualitatively suggest that $\theta_i(t)$ is described by a narrow, 
approximately normal distribution at small times and an increasingly broad, approximately uniform distribution at large times. 
$\theta_i(t)$ is only defined modulo $2\pi$ and can be described as a circular variable defined on the interval 
$-\pi < \theta_i \leq \pi$. 
The distribution of $\theta_i(t)$ can therefore be described with angular histograms, as shown in Fig.~\ref{ThAngularHistograms}. 
Again, $\theta_i(t)$ resembles a uniform circular random variable at large times.

A cumulant expansion can be readily constructed for $M_\theta(t)$. 
The mean  phase is given in terms of the characteristic function and cumulants of $\theta_i(t)$ by
\begin{equation}
  \begin{split}
    \avg{e^{i\theta_i(t)}} = \Phi_{\theta(t)}(1) = \exp\left[\sum_{n=0}^\infty \frac{i^n}{n!}\kappa_n(\theta_i(t))\right]
    \ \ \ ,
  \end{split}
  \label{cumulantTh}
\end{equation}
and  the appropriate cumulant expansion for $M_\theta(t)$ is therefore, 
using Eq.~(\ref{EMThdef}),
\begin{equation}
  \begin{split}
    M_\theta(t) = \sum_{n=0}^\infty \frac{i^n}{n!}\left[ \kappa_n(\theta_i(t)) - \kappa_n(\theta_i(t+1)) \right]
    \ \ \  .
  \end{split}
  \label{cumulantThEM}
\end{equation}
Factors of $i^n$ dictate that a linearly increasing variance of $\theta_i(t)$ makes a positive contribution to $M_\theta(t)$, 
in contradistinction to the slight negative contribution to $M_R(t)$ made by linearly increasing variance of $R_i(t)$. 
Since the mean of $\theta_i(t)$ necessarily vanishes, the variance of $\theta_i(t)$ makes the dominant contribution 
to Eq.~\eqref{cumulantThEM} for approximately normally distributed $\theta_i(t)$. 
For this contribution to be positive, the variance of $\theta_i(t)$ must increase, indicating that  
$\theta_i(t)$ has a StN problem. 
For the case of approximately normally distributed  $\theta_i(t)$, non-zero $M_\theta$  requires a StN problem for the  phase.

\begin{figure}[!ht]
  \centering
  \includegraphics[width=\columnwidth]{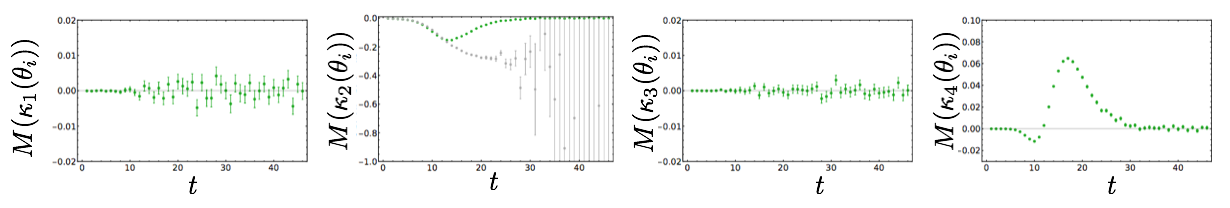}
  \caption{
  Contributions from the first four terms in  the cumulant expansion of Eq.~\eqref{cumulantEM}. 
  The variance, shown second from left, is expected to provide the dominant contribution if a truncation of 
  Eq.~\eqref{cumulantEM} is reliable. 
  Standard estimates of $M_\theta(t)$ from Eq.~\eqref{EMThdef} are shown as the gray points, alongside the cumulant contribution 
  (green points) in the second from left panel. 
  Other panels only show cumulant contributions (green points).
  }
  \label{ThCumulantEM}
\end{figure}

Contributions to Eq.~\eqref{cumulantThEM} from the first four cumulants of $\theta_i(t)$ are shown in Fig.~\ref{ThCumulantEM}. 
Contributions from odd cumulants are consistent with zero, as expected by $\theta_i(t)\rightarrow -\theta_i(t)$ symmetry. 
The variance provides the dominant contribution to $M_\theta(t)$ at small and intermediate times, and is indistinguishable 
from the total $M_\theta(t)$ calculated using the standard effective mass estimator for $t \lesssim 15$. 
Towards to end of the golden window $15 \lesssim t \lesssim 25$, the variance contribution to the effective mass 
begins to decrease. At very large times $t \gtrsim 30$ contributions to $M_\theta(t)$ from the variance are consistent with zero. 
The fourth cumulant makes smaller but statistically significant contributions to $M_\theta(t)$ at intermediate times. 
Contributions from the fourth cumulant also decrease and are consistent with zero at large times. 
The vanishing of these contributions results from the distribution becoming uniform at large times, and time independent as a
consequence.
These observations signal a breakdown in the cumulant expansion at large times $t \gtrsim 25$ where contributions 
from the variance do not approximate standard estimates of $M_\theta(t)$. Notably, the breakdown of the cumulant 
expansion at $t\gtrsim 25$ coincides with plateaus to uniform distribution cumulants in Fig.~\ref{ThCumulants} and 
with the onset of the noise region discussed in Sec.~\ref{sec:decomposition}.

Observations of these unexpected behaviors of $\theta_i(t)$ in the noise region hint at more fundamental issues with the 
statistical description of $\theta_i(t)$ used above. A sufficiently localized probability distribution of a circular random variable 
peaked far from the boundaries of $-\pi <\theta_i(t) \leq \pi$ can be reliably approximated as a standard probability distribution 
of a linear random variable defined on the real line. For broad distributions of a circular variable, the effects of a finite domain 
with periodic boundary conditions cannot be ignored. While circular random variables are not commonly encountered in 
quantum field theory, they arise in many scientific contexts, 
most notably in astronomy, biology, geography, geology, meteorology and oceanography. 
Familiarity with circular statistics is not assumed here, and a few basic results relevant for understanding the statistical 
properties of $\theta_i(t)$ will be reviewed  without proof. Further details can be found in 
Refs.~\cite{Fisher:1995,Mardia:2009,Borradaile:2003} and references therein.

A generic circular random variable $\theta_i$ can be described by two linear random variables $\cos(\theta_i)$ and $\sin(\theta_i)$ 
with support on the line interval $[-1, 1]$ where periodic boundary conditions are not imposed. 
It is the periodic identification of $\theta_i = \pm \pi$ that makes sample moments poor estimators of the distribution of $\theta_i$ and,
 in particular, allows the sample mean of a distribution symmetrically peaked about $\theta_i = \pm \pi$ to be 
 opposite the actual location of peak probability. Parameter estimation for circular distributions can be straightforwardly 
 performed using trigonometric moments of $\cos(\theta_i)$ and $\sin(\theta_i)$. 
 For an ensemble of $N$ random angles $\theta_i$, the first trigonometric moments are defined by the sample averages,
\begin{equation}
  \bar{\cal C} = \frac{1}{N}\sum_i \cos(\theta_i), \hspace{20pt} \bar{\cal S} = \frac{1}{N}\sum_i \sin(\theta_i)
  \ \ \ .
  \label{trigmeandef}
\end{equation}
Higher trigonometric moments can be defined analogously but will not be needed here. 
The average angle can be defined in terms of the mean two-dimensional vector $(\bar{\cal C},\;\bar{\cal S})$ as
\begin{equation}
  \bar{\theta} = \text{arg}\left( \bar{\cal C} + i \bar{\cal S} \right)
  \ \ \  .
  \label{thetabardef}
\end{equation}
A standard measure of a circular distribution's width is given in terms of trigonometric moments as
\begin{equation}
  \bar{\rho}^2 = \bar{\cal C}^2 + \bar{\cal S}^2
  \ \ \ 
  \label{Rdef}
\end{equation}
where $\bar{\rho}$ should be viewed as a measure of the concentration of a circular distribution. 
Smaller $\bar{\rho}$ corresponds to a broader, more uniform distribution, while larger $\bar{\rho}$ corresponds to a more localized distribution.

One way of defining statistical distributions of circular random variables is by ``wrapping'' distributions for linear random 
variables around the unit circle. The probability of a circular random variable equaling some value in $-\pi < \theta \leq \pi$ 
is equal to the sum of the probabilities of the linear random variable equaling any value that is equivalent to $\theta$ modulo $2\pi$. 
Applying this prescription to a normally distributed linear random variable gives the wrapped normal distribution
\begin{equation}
  \begin{split}
  \mathcal{P}_{WN}(\theta_i;\mu,\sigma) &= 
  \frac{1}{\sqrt{2\pi}\sigma}\sum_{k=-\infty}^\infty \exp\left[ -\frac{(\theta_i - \mu + 2\pi k)^2}{2\sigma^2} \right]
  \ =\  \frac{1}{2\pi}\sum_{n=-\infty}^\infty e^{in(\theta_i - \mu) -\sigma^2 n^2/2}
  \ \ \  ,
  \label{WNdef}
\end{split}
\end{equation}
where the second form follows from the Poisson summation formula. 
Wrapped distributions share the same characteristic functions as their unwrapped counterparts, 
and the second expression above can be derived as a discrete Fourier transform of a normal characteristic function. 
The second sum above can also be compactly represented in terms of elliptic-$\vartheta$ functions. 
For $\sigma^2 \lesssim 1$ the wrapped normal distribution qualitatively resembles a normal distribution, 
but for $\sigma^2 \gtrsim 1$ the effects of wrapping obscure the localized peak. 
As $\sigma^2 \rightarrow \infty$, the wrapped normal distribution becomes a uniform distribution on $(-\pi,\pi]$. 
Arbitrary trigonometric moments and therefore the characteristic function of the wrapped normal distribution are given by 
\begin{equation}
   \avg{e^{i n \theta_i}}_{WN} = e^{in\mu - n^2\sigma^2/2}
   \ \ \ \  .
  \label{WNmoments}
\end{equation}
Parameter estimation in fitting a wrapped normal distribution to LQCD results for $\theta_i(t)$ 
can be readily performed by relating $\bar{\theta}$ and $\bar{\rho}$ above to these trigonometric moments as
\begin{equation}
  \mu = \bar{\theta}
  \qquad {\rm and }\qquad
 e^{-\sigma^2} = \bar{\rho}^2
 \ \ \  .
  \label{WNest}
\end{equation}
Note that Eq.~\eqref{WNest} holds only in the limit of infinite statistics. 
Estimates for the average of a wrapped normal distribution 
are consistent with zero at all times, as expected. 
Wrapped normal probability distribution functions with $\sigma^2(\theta_i(t))$ determined from Eq.~\eqref{WNest} 
are shown with the histograms of Fig.~\ref{ThHistograms} and provide a good fit to the data at all times.

The appearance of a uniform distribution at large times is consistent with the heuristic argument that the logarithm of a 
correlation function should be described by a stable distribution. The uniform distribution is a stable distribution for 
circular random variables, and in fact is the only stable circular distribution~\cite{Mardia:2009}. The distribution describing 
a sum of many linear random variables broadens as the number of summands is increased, and the same is true of circular 
random variables. A theorem of Poincar{\'e} proves that as the width of any circular distribution is increased without bound, 
the distribution will approach a uniform distribution. One therefore expects that the sum of many well-localized 
circular random variables might initially tend towards a narrow wrapped normal distribution while boundary effects are negligible. 
Eventually as more terms are added to the sum this wrapped normal distribution will broaden and approach a uniform distribution. 
This intuitive picture appears consistent with the time evolution of $\theta_i(t)$  shown in 
Figs.~\ref{ThHistograms},~\ref{ThAngularHistograms}.

\begin{figure}[!ht]
  \centering
  \includegraphics[width=\columnwidth]{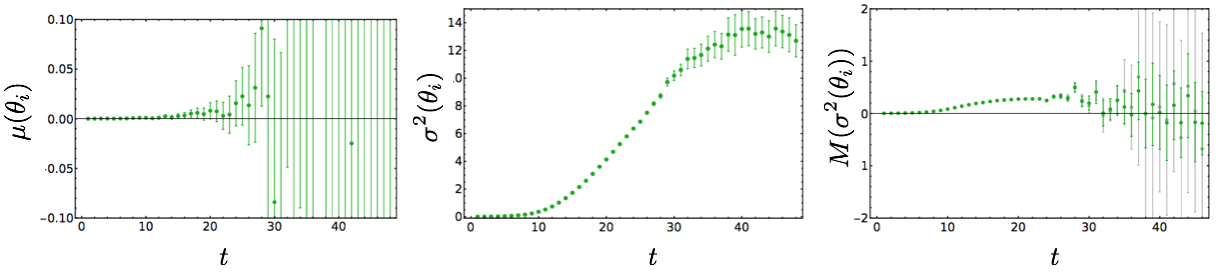}
  \caption{
    The left panel shows estimates of the wrapped normal mean $\mu(\theta_i(t))$ calculated from Eq.~\eqref{WNest} as a function of time.
  The center panel shows analagous estimates of the wrapped normal variance, $\sigma^2(\theta_i(t))$.
  The right panel shows the wrapped normal effective mass, $M_\theta^{WN}(t)$, 
  defined in Eq.~\eqref{mthetaWN}  (green points) 
  along with the standard complex phase effective mass $M_\theta(t)$ 
 defined in Eq.~\eqref{cumulantThEM}  (gray points).
 }
  \label{ThWN}
\end{figure}
The wrapped normal variance estimates for $\theta_i(t)$ that are shown in Fig.~\ref{ThWN} require further discussion. 
At intermediate times, the wrapped normal variance calculated from Eq.~\eqref{WNest} rises linearly with a slope consistent with $M_N - \frac{3}{2}m_\pi$. This is not surprising because assuming an exactly wrapped normal $\theta_i(t)$, $M_\theta(t)$ becomes
\begin{equation}
  \begin{split}
    M_\theta^{WN}(t) = \ln\left[\frac{\avg{e^{i\theta_i(t)}}_{WN}}{\avg{e^{i\theta_i(t+1)}}_{WN}}\right] = -\frac{1}{2}\left[\sigma^2(\theta_i(t)) - \sigma^2(\theta_i(t+1))\right]
    \ \ \  .
  \end{split}
  \label{mthetaWN}
\end{equation}
Eq.~\eqref{mthetaWN} resembles the first non-zero term in the cumulant expansion given in Eq.~\eqref{cumulantThEM} 
adapted for circular random variables. 
Results for $M_\theta^{WN}(t)$ are also shown in Fig.~\ref{ThWN}, where it is seen that $M_\theta^{WN}(t)$ is indistinguishable 
from $M_\theta(t)$ at small and intermediate times. In the noise region, both $M_\theta^{WN}(t)$ and standard estimates for 
$M_\theta(t)$ are consistent with zero. $M_\theta(t)$ has smaller variance than $M_\theta(t)$ in the noise region, 
but this large-time noise is the only visible signal of deviation between the two. This is not surprising, because 
$M_\theta^{WN}(t)$ is actually identical to $M_\theta(t)$ when $\bar{\cal S}(\theta(t)) = 0$. Since $\bar{\cal S}(\theta_i(t))$ 
vanishes in the infinite statistics limit by $\theta_i(t) \rightarrow -\theta_i(t)$ symmetry, $M_\theta^{WN}(t)$ must 
agree with $M_\theta(t)$ up to statistical noise. At large times $t\gtrsim 30$, the wrapped normal variance shown in 
Fig.~\ref{ThWN}  becomes roughly constant up to sizable fluctuations. 
The region where $\sigma^2(\theta_i(t))$ stops increasing coincides with the noise region previously identified.

\begin{figure}[!ht]
  \centering
  \includegraphics[width=.5\columnwidth]{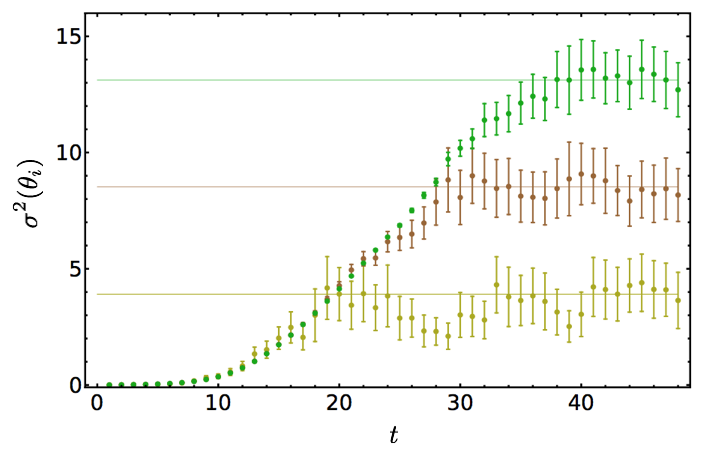}
  \caption{
  Wrapped normal variance of the phase $\sigma^2(\theta_i(t))$ for statistical ensembles of various sizes. 
  Results for an ensemble of $N=50$ nucleon correlation functions are shown in yellow, 
  $N=5,000$ in brown, and $N=500,000$ in green. 
  Lines of each color are also shown at $\sigma^2(\theta_i(t)) = \ln(N)$. 
  Above the relevant line, Eq.~\eqref{Nscaling} is violated for each ensemble and measurements of $\sigma^2(\theta_i(t))$ 
  are expected to be roughly equal to $\ln(N)$ instead of the underlying physical value of $\sigma^2(\theta_i(t))$. 
  Estimates of $\sigma^2(\theta_i(t))$ reaching these lines marks the beginning of the noise region defined by 
  violations of Eq.~\eqref{Nscaling} for each ensemble.}
  \label{ThWNAll}
\end{figure}

The time at which  the noise region begins depends on the size of the statistical ensemble $N$. 
Figure~\ref{ThWNAll} shows estimates of $\sigma^2(\theta_i(t))$ from Eq.~\eqref{WNest} for statistical ensemble sizes $N=50,\;5,000,\;500,000$ varying across four orders of magnitude. 
The time of the onset of the noise region  varies logarithmically as $t \sim 20,\;27,\;35$. 
The constant noise region value of $\sigma^2(\theta_i(t))$ is also seen to vary logarithmically with $N$. 
Equality of $M_\theta^{WN}(t)$ and $M_\theta(t)$ up to statistical fluctuations shows that $M_\theta(t)$ must be 
consistent with zero in the noise region. 
Since corrections to $M(t) \approx M_\theta(t) + M_R(t)$ from magnitude-phase correlations appear small at all times, 
it is reasonable to conclude that standard estimators for the nucleon effective mass are systematically biased in the noise 
region and that exponentially large increases in statistics are required to delay the onset of the noise region.

Besides these empirical observations, the inevitable existence and exponential cost of delaying the noise region 
can be understood from general arguments of circular statistics. 
The expected value of the sample concentration $\bar{\rho}^2$ can be calculated by 
applying Eq.~\eqref{WNmoments} to an ensemble of independent wrapped normal random variables 
$\theta_i$ in Eq.~\eqref{Rdef}. The result shows that $\bar{\rho}^2$ is a biased estimate of $e^{-\sigma^2}$, 
and that the appropriate unbiased estimator is~\cite{Fisher:1995,Mardia:2009}
\begin{equation}
  e^{-\sigma^2} = \frac{N}{N-1}\left( \bar{\rho}^2 - \frac{1}{N} \right).
  \label{Redef}
\end{equation}
For $\bar{\rho}^2 < 1/N$, Eq.~\eqref{Redef} would lead to an imaginary estimate for $\sigma^2$ and therefore 
no reliable unbiased estimate can be extracted. 
A similar calculation shows that the expected variance of $\bar{\rho}^2$ is
\begin{equation}
  \begin{split}
    \text{Var}(\bar{\rho}^2) 
    = 
    \frac{N-1}{N^3}
    \left(1 - e^{-\sigma^2} \right)^2
    \left[\  \left(1 - e^{-\sigma^2} \right)^2 + 2 N e^{-\sigma^2} \ \right]
    \ \ \ .
  \end{split}\label{VarRbar}
\end{equation}
In the limit of an infinitely broad distribution, all circular distributions tend towards uniform and the variance 
of $\bar{\rho}^2$ is set by the $\sigma^2\rightarrow\infty$ limit of Eq.~\eqref{VarRbar} regardless of the form of 
the true underlying distribution. When analyzing any very broad circular distribution, measurements of $\bar{\rho}^2$ 
will therefore include fluctuations on the order of $1/N$. For $e^{-\sigma^2} < 1/N$, the expected error from finite 
sample size effects in statistical inference based on $\bar{\rho}^2$ is therefore larger than the signal to be measured. 
In this regime $\bar{\rho}^2$ has both systematic bias and expected statistical errors that are larger than the 
value $e^{-\sigma^2}$ that $\bar{\rho}^2$ is supposed to estimate. $\bar{\rho}^2$ cannot provide accurate 
estimates of $e^{-\sigma^2}$ in this regime.

Inability to perform statistical inference in the regime $e^{-\sigma^2} < 1/N$ matters for the nucleon correlation 
function because $e^{-\sigma^2(\theta_i(t))} = \bar{\rho}^2(\theta_i(t))= \avg{\cos(\theta_i(t))}^2$ and therefore $e^{-\sigma^2(\theta_i(t))}$ 
decreases exponentially with time. 
At large times there will necessarily be a noise region where $e^{-\sigma^2(\theta_i(t))} < 1/N$ is reached and $\bar{\rho}^2(\theta_i(t))$ 
is not a reliable estimator.
Keeping $e^{-\sigma^2(\theta_i(t))}$ larger than the bias and expected fluctuations of $\bar{\rho}^2(\theta_i(t))$ requires
\begin{equation}
  N > e^{\sigma^2(\theta_i(t))} \sim e^{2\left( M_N - \frac{3}{2}m_\pi \right)t}
  \ \ \ .
  \label{Nscaling}
\end{equation}
Eq.~\eqref{Nscaling} demonstrates that exponential increases in statistics are required to delay the time where statistical uncertainties 
and systematic bias dominate physical results estimated from $\bar{\rho}^2(\theta_i(t))$. 
Formally, the noise region can be defined as the region where Eq.~\eqref{Nscaling} is violated. 
Lines at $\sigma^2(\theta_i(t)) = \ln N$ are shown on Fig.~\ref{ThWNAll} for the ensembles with $N=50,\;5,000,\;500,000$ shown. 
By this definition, the noise region formally begins once $\sigma^2(\theta_i(t))$
 (extrapolated from reliable estimates in the golden window) crosses above the appropriate line. 
 Excellent agreement can be seen between this definition and the above empirical characterizations of the 
 noise region based on constant $\sigma^2(\theta_i(t))$ and unreliable effective mass estimates with constant errors.

Breakdown of statistical inference for sufficiently broad distributions is a general feature of circular distributions. 
Fisher notes that circular distributions are distinct from more familiar linear distributions in that 
``formal statistical analysis cannot proceed'' for sufficiently broad distributions~\cite{Fisher:1995}. 
The arguments above do not rely on the particular form of the wrapped normal model assumed for $\theta_i(t)$, 
and the basic cause for the onset of the noise region for broad $\theta_i(t)$ is that $\bar{\rho}^2$ has an uncertainty of order 
$1/N$ for any broad circular distribution that begins approaching a uniform distribution.~\footnote{
One may wonder whether there is a more optimal estimator than $\bar{\rho}^2$ that could reliably calculate 
the width of broad circular distributions with smaller variance. 
While this possibility cannot be discarded in general, it is interesting to note that it can be in one model. 
The most studied distribution in one-dimensional circular statistics is the von Mises distribution, which has a 
simpler analytic form than the wrapped normal distribution. 
The von Mises distribution is also normally distributed in the limit of a narrow distribution, uniform in the limit of a broad distribution, 
and in general a close approximation but not identical to the wrapped normal distribution. 
Von Mises distributions provide fits of comparable qualitative quality to $\theta_i(t)$ as wrapped normal distributions. 
For the von Mises distribution, $\frac{N}{N-1}\left(\bar{\rho}^2 - \frac{1}{N}\right)$ is an unbiased maximum likelihood 
estimator related to the width. By the Cram{\'e}r-Rao inequality, a lower mean-squared error cannot be achieved 
if $\theta_i(t)$ is von Mises. 
Particularly in the limit of a broad distribution where all circular distributions tend towards uniform, 
it would be very surprising if an estimator could be found that satisfied this bound for the von Mises 
case but could reliably estimate the width of $\theta_i(t)$ in the noise region if a different underlying 
distribution is assumed.
}
Analogs of Eq.~\eqref{Nscaling} can be expected to apply to statistical estimation of the mean of any complex correlation function. 
As long as the asymptotic value of $M_\theta$ is known, Eq.~\eqref{Nscaling} and analogs for other complex correlation 
functions can be used to estimate the required statistical ensemble size necessary to reliably estimate the mean 
correlation function up to a desired time $t$.

\begin{figure}[!ht]
  \centering
  \includegraphics[width=\columnwidth]{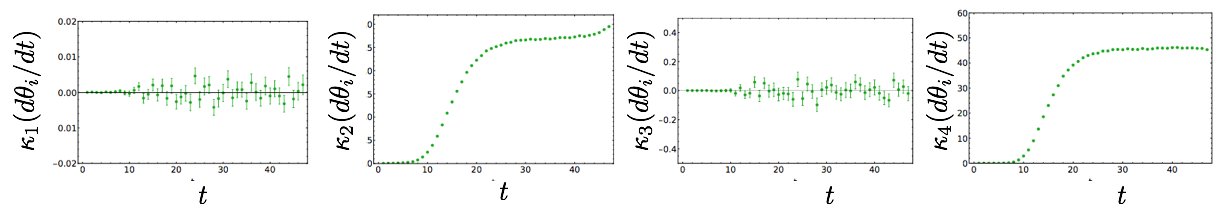}
  \caption{The first four cumulants of $\frac{d\theta_i}{dt}$.}
  \label{dThdtCumulants}
\end{figure}

\begin{figure}[!ht]
  \centering
  \includegraphics[width=\columnwidth]{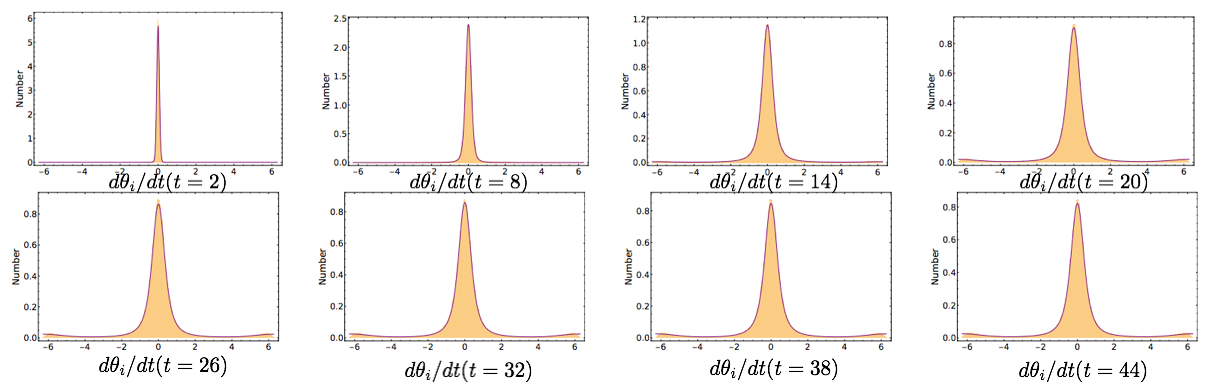}
  \caption{
  Histograms of $\frac{d\theta_i}{dt}$ with fits to  a wrapped stable mixture distribution shown as the purple curves. 
  See the main text for details.
  }
  \label{dThdtHistograms}
\end{figure}

\begin{figure}[!ht]
  \centering
  \includegraphics[width=\columnwidth]{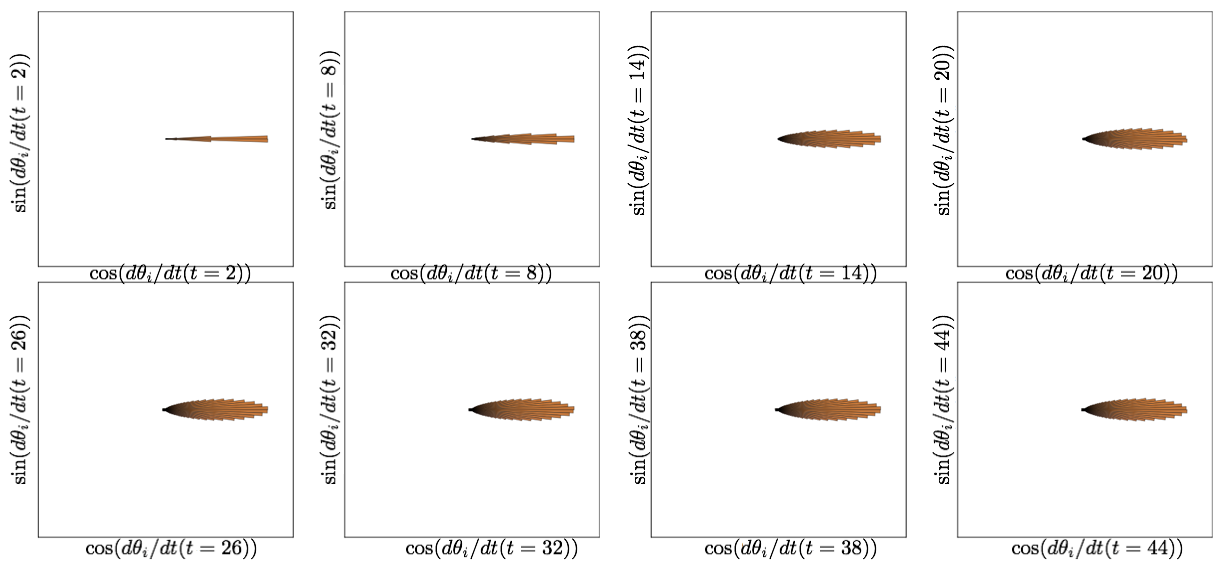}
  \caption{
  Angular histograms of $\frac{d\theta_i}{dt}$. 
  Since $\frac{d\theta_i}{dt}$ is defined on $-2\pi < \frac{d\theta_i}{dt} \leq 2\pi$, 
  normalizations are such that $\frac{1}{2}\frac{d\theta_i}{dt}$ is mapped to the unit circle in analogy to 
  Fig.~\ref{ThAngularHistograms}.}
  \label{dThdtAngularHistograms}
\end{figure}

The pathological features of the large-time distribution of $\theta_i$ are not shared by $\frac{d\theta_i}{dt}$. 
As with the log-magnitude, 
it is useful to 
define general finite differences,
\begin{equation}
  \begin{split}
    \Delta\theta_i(t,\Delta t) = \theta_i(t) - \theta_i(t-\Delta t)
    \ \ \ \  ,
  \end{split}
  \label{DeltaThdef}
\end{equation}
and a discrete (lattice) time derivative,
\begin{equation}
  \begin{split}
    \frac{d\theta_i}{dt} = \Delta\theta_i(t, 1)
    \ \ \ \  .
  \end{split}
  \label{dThdtdef}
\end{equation}
The sample cumulants of $\frac{d\theta_i}{dt}$ are shown in Fig.~\ref{dThdtCumulants}, 
histograms of $\frac{d\theta_i}{dt}$ are shown in Fig.~\ref{dThdtHistograms}, and angular histograms are 
shown in Fig.~\ref{dThdtAngularHistograms}. 
Much like $\frac{dR_i}{dt}$, $\frac{d\theta_i}{dt}$ appears to have a time independent distribution at large times. While $\frac{d\theta_i}{dt}$ is a circular random variable, it's distribution is still well-localized at large times and can be clearly visually distinguished from a uniform distribution. This suggests that statistical inference of $\frac{d\theta_i}{dt}$ should be reliable in the noise region.

\begin{figure}[!ht]
  \centering
  \includegraphics[width=\columnwidth]{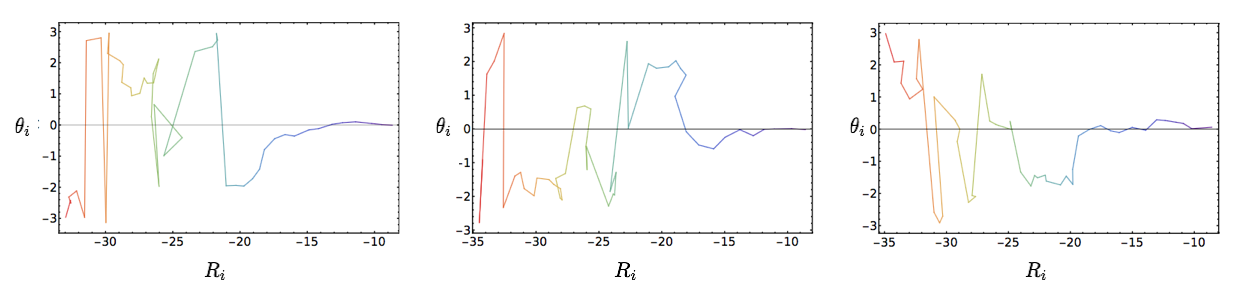}
  \caption{
  Time series showing $R_i(t)$ on the horizontal axis and $\theta_i(t)$ on the vertical axis for three 
  individual nucleon correlation functions,
  where the color of the line shows the time evolution from violet at $t=0$ to red at $t=48$. 
  The evolution of $R_i(t)$ shows a clear drift towards increasingly negative $R_i(t)$. 
  Some large jumps where $\theta_i(t)$ changes by nearly $\pm 2\pi$  
  correspond to crossing the branch cut in $\theta_i(t)$. 
  There are also sizable jumps where $\theta_i(t)$ changes by nearly $\pm\pi$ which likely do not 
 correspond to crossing a branch cut. 
  }
  \label{Tracks}
\end{figure}

Like $\frac{dR_i}{dt}$, $\frac{d\theta_i}{dt}$ shows evidence of heavy tails. 
The time evolution of $R_i(t)$ and $\theta_i(t)$ for three (randomly selected)
correlation functions are shown in Fig.~\ref{Tracks}, 
and exhibit large jumps in both $R_i(t)$ and $\theta_i(t)$ more characteristic of L{\'e}vy flights than Brownian motion,
leading us to consider stable distributions once again.
Wrapped stable distributions can be constructed analogously to wrapped normal distributions as
\begin{equation}
  \begin{split}
    \mathcal{P}_{WS}(\theta_i; \alpha, \beta, \mu, \gamma) &= 
    \sum_{k=-\infty}^\infty \mathcal{P}_S(\theta_i + 2\pi k; \alpha, \beta, \mu, \gamma) \\
    &= \frac{1}{2\pi}\sum_{n=-\infty}^\infty \exp\left( i\mu n - |\gamma n|^\alpha\left[ 1 - i\beta\frac{n}{|n|}\tan(\pi \alpha/2) \right] \right)
    \ \ \   ,
  \end{split}
  \label{PWSdef}
\end{equation}
where, as in Eq.~\eqref{PhiSdef}, 
$\tan(\pi \alpha/2)$ should be replaced by $-\frac{2}{\pi}\ln|n|$ for $\alpha = 1$. 
This wrapped stable distribution is still not appropriate to describe $\frac{d\theta_i}{dt}$ for two reasons. 
First, $\frac{d\theta_i}{dt}$ describes a difference of angles and so is defined on a periodic domain 
$-2\pi < \frac{d\theta_i}{dt} \leq 2\pi$. 
This is trivially accounted for by replacing $2\pi$ by $4\pi$ in Eq.~\eqref{PWSdef}. 
Second, $\theta_i(t)$ is determined from a complex logarithm of $C_i(t)$ with a branch cut placed at $\pm \pi$. 
Whenever $\theta_i(t)$ makes a small jump across this branch cut, $\frac{d\theta_i}{dt}$ will be measured to be around $2\pi$ 
even though the distance traveled by $\theta_i(t)$ along its full Riemann surface is much smaller. 
This behavior results in the small secondary peaks near $\frac{d\theta_i}{dt} = \pm 2\pi$ visible in Fig.~\ref{dThdtHistograms}. 
This can be accommodated by fitting $\frac{d\theta_i}{dt}$ to a mixture of wrapped stable distributions 
peaked at zero and $2\pi$. Since $\theta_i(t)\rightarrow-\theta_i(t)$ symmetry demands that both of these 
distributions are symmetric, a probability distribution able to accommodate all observed features of 
$\frac{d\theta_i}{dt}$ is given by the wrapped stable mixture distribution
\begin{equation}
  \begin{split}
    \tilde{\mathcal{P}}_{WS}(\theta_i; \alpha_1, \alpha_2, \gamma_1, \gamma_2, f) &= 
    \frac{1}{4\pi}\left[ 1 + 2\sum_{n=1}^\infty (1-f)e^{-|\gamma_1 n|^{\alpha_1}}\cos(n\theta_i) 
    + f e^{-|\gamma_2 n|^{\alpha_2}}\cos(n(\theta_i - 2\pi)) \right]
    \ \ \  ,
  \end{split}
  \label{PWSMixdef}
\end{equation}
where $f$ represents the fraction of $\frac{d\theta_i}{dt}$ data in the secondary peaks at $\frac{d\theta_i}{dt} = \pm 2\pi$ 
representing branch cut crossings. Fits of $\frac{d\theta_i}{dt}$ to this wrapped stable mixture model performed with 
maximum likelihood estimation are shown in Fig.~\ref{dThdtHistograms} and are in good qualitative agreement with the LQCD results.

If the widths of the main and secondary peaks in $\frac{d\theta_i}{dt}$  were sufficiently narrow, it would be possible to 
unambiguously associate each $\frac{d\theta_i}{dt}$ measurement with one peak or the other and ``unwrap'' the 
trajectory of $\theta_i(t)$ across its full Riemann surface by adding $\pm 2\pi$ to measured values of $\frac{d\theta_i}{dt}$ 
whenever the branch cut in $\theta_i(t)$ is crossed. 
This should become increasingly feasible as the continuum limit is approached. 
However, the presence of heavy tails in the $\frac{d\theta_i}{dt}$ primary peak prevent unambiguous identification of branch cut 
crossings in the LQCD correlation functions considered here. 
Due to the  power-law decay of the primary peak, there is no clear separation visible between the main and secondary peaks, 
and in particular, points near $\frac{d\theta_i}{dt} = \pm \pi$ cannot be unambiguously identified with one peak or another.

\begin{figure}[!ht]
  \centering
  \includegraphics[width=\columnwidth]{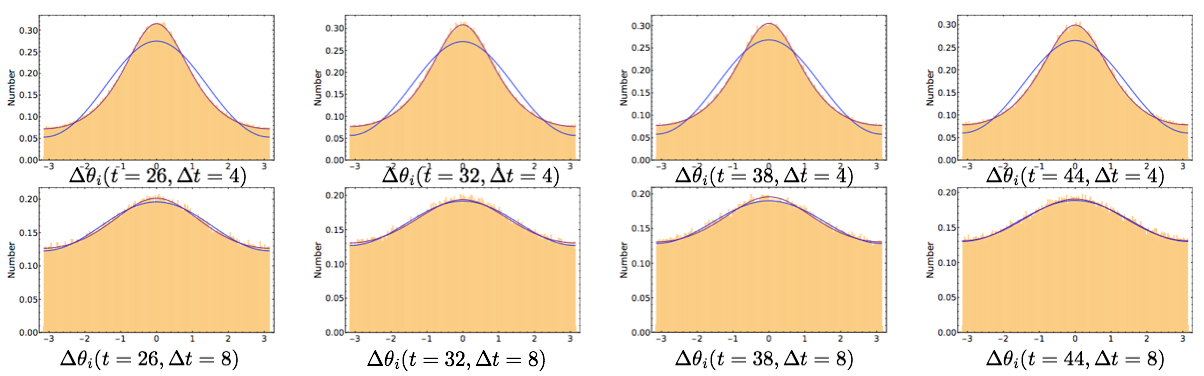}
  \caption{
  Histograms of  $\widetilde{\Delta \theta_i}$ along with fits to wrapped normal distributions in blue and wrapped stable distributions in purple.
  }
  \label{DeltaThHistograms}
\end{figure}

\begin{figure}[!ht]
  \centering
  \includegraphics[width=\columnwidth]{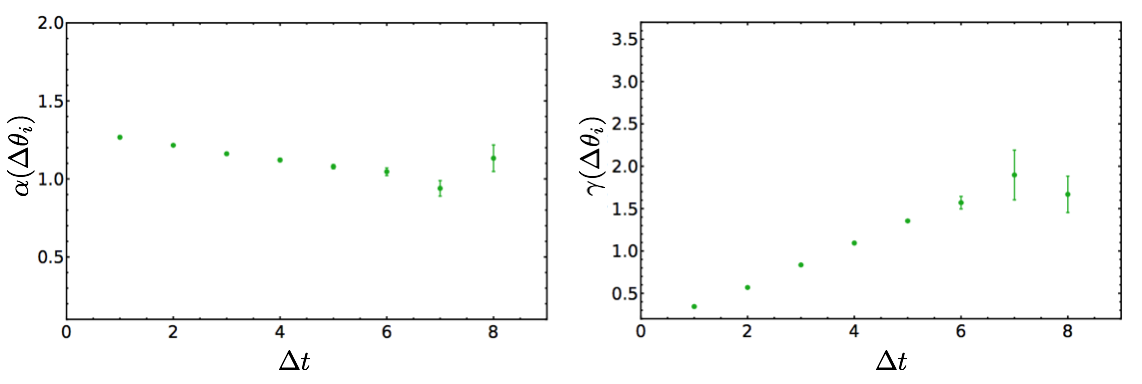}
  \caption{Maximum likelihood estimates for the wrapped stable index of stability $\alpha\left( \widetilde{\Delta \theta_i} \right)$, left, and width $\gamma\left( \widetilde{\Delta \theta_i} \right)$, right extracted from the large-time plateau region as functions of $\Delta t$.
}
  \label{DeltaThStable}
\end{figure}

For descriptive analysis of $\frac{d\theta_i}{dt}$, it is useful to shift the secondary peak to the origin by defining
\begin{equation}
  \begin{split}
    \widetilde{\Delta \theta_i} = \text{Mod}\left( \Delta \theta_i + \pi, 2\pi \right) - \pi
    \ \ \ \ .
  \end{split}
  \label{DeltaThTdef}
\end{equation}
$\widetilde{\Delta \theta_i}$ is well-described by the wrapped stable distribution of Eq.~\eqref{PWSdef}. 
Histograms of the large-time behavior of $\widetilde{\Delta \theta_i}$ are shown in Fig.~\ref{DeltaThHistograms} for $\Delta t = 4,\; 8$ 
and fits of the index of stability of $\widetilde{\Delta \theta_i}$ are shown in Fig.~\ref{DeltaThStable}. 
The large-time distribution of $\widetilde{\Delta \theta_i}(t,\Delta t)$ appears time independent for all $\Delta t$. 
Heavy tails are visible at all times, even as $\Delta t$ becomes large. The large $\Delta t$ behavior visible here is consistent with a wrapped Cauchy distribution.
The estimated index of stability of $\widetilde{\Delta\theta_i}$ differs significantly from that of $\Delta R$, and for $\Delta t = 1$, the large-time behavior is found to have
\begin{equation}
  \begin{split}
    \alpha\left( \widetilde{\Delta\theta_i}(t\rightarrow\infty, \Delta t \sim 0.12\text{ fm}) \right) \rightarrow 1.267(4)(1).
    \ \ \ \  .
  \end{split}
  \label{alpha1}
\end{equation}
This result is consistent with maximum likelihood estimates of $\alpha_1\left( \frac{d\theta_i}{dt} \right)$ in the 
wrapped stable mixture model of Eq.~\eqref{PWSMixdef}. 
$\alpha_2\left( \frac{d\theta_i}{dt} \right)$, associated with the peak shifted from $\theta_i=\pm\pi$
in the wrapped stable mixture model, 
cannot be reliably estimated from the  available LQCD correlation functions. 
The continuum limit index of stability of $\frac{d\theta_i}{dt}$ cannot be determined without additional LQCD studies 
at finer lattice spacings.

As seen in Fig.~\ref{DeltaThStable}, 
the large-time width of $\widetilde{\Delta \theta_i}(t, \Delta t)$ increases with increasing $\Delta t$. 
This behavior is shared by $\Delta \theta_i(t, \Delta t)$. 
In accordance with the observations above that the wrapped normal variance of $\theta_i(t)$ increases linearly with $t$, 
the constant large-time wrapped normal variance of $\Delta\theta_i(t, \Delta t)$ increases linearly with $\Delta t$. 
This is consistent with a pciture of $\Delta\theta_i(t,\Delta t)$ as the sum of $\Delta t$ single time step differences, $\frac{d\theta_i}{dt}$, that make roughly equal contributions to $\Delta \theta_i(t, \Delta t)$.
In accordance with the scaling $\sigma^2(\theta_i(t)) \sim (M_N - \frac{3}{2}m_\pi)t$ discussed previously, 
this linear scaling gives $\sigma^2(\Delta \theta_i(t, \Delta t)) \sim 2(M_N - \frac{3}{2}m_\pi)\Delta t$.

We summarize our observations on the  phase of $C(t)$:
\begin{itemize}
  \item 
  The  phase of the nucleon correlation function is described by an approximately wrapped normal distribution 
  whose width increases with time. At small times the distribution is narrow and resembles a normal distribution. 
  At large times the distribution becomes broad compared to the $2\pi$ range of definition of $\theta_i(t)$ and resembles a uniform distribution. 
  \item 
  The phase effective mass $M_\theta(t)$ appears to plateau to a value close to $M_N - 3/2 m_\pi$. 
  Since $|e^{i\theta_i(t)}|^2 = 1$ is time-independent by construction, this non-zero asymptotic value of 
  $M_\theta$ implies $\theta_i(t)$ has a severe StN problem.
  \item 
  $M_\theta(t)$ can be determined from the time derivative of the wrapped normal variance of $\theta_i(t)$ 
  in analogy to the cumulant expansion. The effective mass extracted from growth of the wrapped 
  normal variance is identical to $M_\theta(t)$ up to statistical fluctuations. 
  This leads to scaling of the wrapped normal variance of $\theta_i(t)$ consistent with 
  $\sigma^2(\theta_i(t)) \sim 2(M_N - \frac{3}{2}m_\pi)t$.
  \item 
  Standard estimators for the wrapped normal variance have a systematic bias and for a sufficiently broad distribution the 
  minimum expected statistical uncertainty is set by finite sample size $1/N$ effects. 
  Once the wrapped normal variance becomes larger than $\ln N$, finite sample size fluctuations 
  become larger than the signal required to extract $M_\theta(t)$. 
  Since the width of $\theta_i(t)$ increases with time, a region where finite sample size errors prevent reliable extractions of $M_\theta(t)$ will inevitably occur at sufficiently large times. This is the noise region empirically identified above. Standard effective mass estimates are systematically biased in the noise region. Exponentially large increases in statistics are necessary to delay the onset of the noise region. 
  \item 
  Finite differences, $\Delta \theta_i(t,\Delta t)$, are described by time-independent distributions at large times. 
  $\Delta \theta_i$ is heavy-tailed for all $\Delta t$ considered here, and  $\frac{d\theta_i}{dt}$ 
  is well-described by a wrapped stable mixture distribution. 
  Further studies will be needed to understand the continuum limit of the  index of stability of $\frac{d\theta_i}{dt}$. 
\end{itemize}

\section{An Improved Estimator}\label{sec:estimator}

The proceeding observations suggest that difficulties in statistical analysis of nucleon correlation functions 
arise from difficulties in statistical inference of $\theta_i(t)$. The same exponentially hard StN and noise region 
problems obstruct large-time estimation of the wrapped normal variance of $\theta_i(t)$ and of $M(t)$. 
Conversely, the width of $\Delta \theta_i(t,\Delta t)$ distributions does not increase with time, and there is 
no StN problem impeding statistical inference of $\Delta \theta_i(t,\Delta t)$. 
This suggests that it would be preferable to construct an effective mass estimator relying on statistical 
inference of $\Delta\theta_i(t,\Delta t)$.

\begin{figure}[!ht]
  \centering
  \includegraphics[width=\columnwidth]{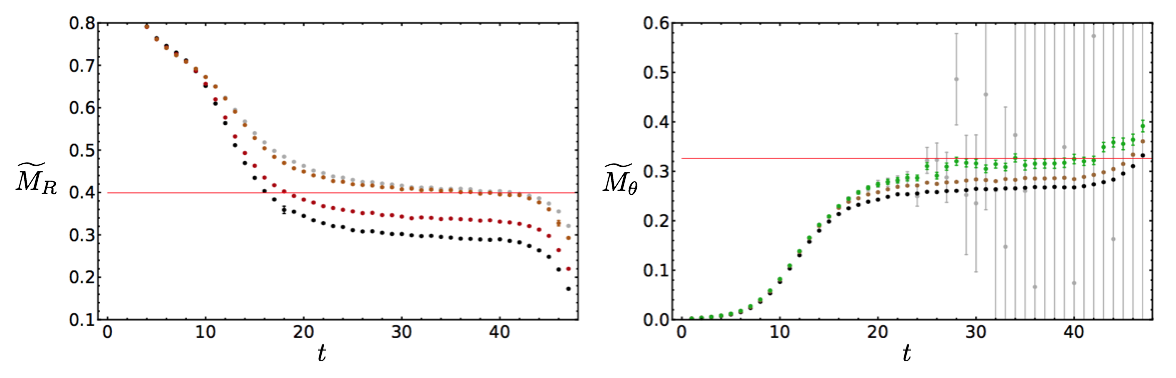}
  \caption{
    Results for the correlation-function-ratio-based estimators $\tilde{M}_R(t, \Delta t)$ and $\tilde{M}_\theta(t, \Delta t)$ with $\Delta t = 1,\;2,\;8$. 
  The left panel shows results for $m_R(t, \Delta t)$ with $\Delta t = 1$ in black, $\Delta t = 2$ in red, and $\Delta t = 8$ in orange. 
  The standard estimator $m_R(t)$ is shown in gray, and a red line is shown for reference at $\frac{3}{2}m_\pi$. 
  The right panel shows results for $m_\theta(t, \Delta t)$ with $\Delta t = 1$ in black, $\Delta t = 2$ in brown, and $\Delta t = 8$ in green. 
  The standard estimator $m_\theta(t)$ is shown in gray and a red line is shown for reference at $M_N - \frac{3}{2}m_\pi$.
  }
  \label{RThWEM}
\end{figure}

First consider the magnitude for simplicity. 
 The mean correlation function magnitude can be expressed in terms of $\Delta R_i$ as
 \begin{equation}
   \begin{split}
     \avg{ e^{R_i(t)} } &= \avg{ \exp\left(R_i(0) + \sum_{t^\prime = 1}^{t} \left.\frac{dR_i}{dt}\right|_{t^\prime} \right)}\\
     &= \avg{ \exp\left(R_i(0) +  \sum_{t^\prime = 1}^{t - \Delta t} \left.\frac{dR_i}{dt}\right|_{t^\prime} \right) \exp\left( \sum_{t^\prime = t- \Delta t + 1}^{t} \left.\frac{dR_i}{dt}\right|_{t^\prime} \right) } \\
     &= \avg{ e^{R_i(0) + \Delta R_i(t - \Delta t, t - \Delta t)}  e^{\Delta R_i(t, \Delta t)}  }
     \ \ \ \   .
   \end{split}
   \label{meanDeltaR}
 \end{equation}
 The last expression above shows that $e^{R_i(t)}$ can be expressed as a product of two factors involving the
 evolution of $R_i(t)$ in the regions $[0, t-\Delta t]$ and $[t-\Delta t, t]$ respectively. 
 Because QCD has a finite correlation length, these two factors should be approximately decorrelated.
 Correlations should only arise from contributions involving points near the boundary at $t- \Delta t$.
 At large times, $t$ can be assumed to be much larger than $\Delta t$ and than any QCD correlation length,
 so boundary effects can be assumed to be negligible for the first region.
 Boundary effects cannot be neglected for the smaller region of length $\Delta t$.
 Treating these boundary effects as a systematic uncertainty allows the correlation function
 to be factorized between the regions $[0, t-\Delta t]$ and $[t-\Delta t, t]$ as
 \begin{equation}
   \begin{split}
     \avg{ e^{R_i(t)} } &= \avg{ e^{R_i(0) + \Delta R_i(t - \Delta t, t - \Delta t)}}  \avg{e^{ \Delta R_i(t, \Delta t) } } 
     \left[ 1   + O\left( e^{-\delta E \Delta t} \right) \right]
     \ \ \ \  .
   \end{split}
   \label{meanDeltaRsplit1}
 \end{equation}
  where $\delta E$ is the smallest energy scale responsible for non-trivial correlations between the factors on the rhs associated with $[0,t-\Delta t]$ and $[t-\Delta t, t]$, and terms suppressed by $e^{-\delta E(t-\Delta t)}$ are neglected.
  If both factors on the rhs of Eq.~\eqref{meanDeltaRsplit1} only receive contributions from the ground state and have single-exponential time evolution,
  then the product of the independently averaged factors on the rhs has the
  same single-exponential behavior as the lhs.
  If excited states make appreciable contributions to either factor on the rhs, then the product of sums of exponentials representing multi-state evolution over $[0,t-\Delta t]$ and $[t-\Delta t, t]$ respectively will not exactly equal the sum of exponentials representing multi-state evolution over $[0,t]$.
  This suggests that $\delta E$ should be set by the gap between the ground state and first excited state with appropriate quantum numbers.\footnote{  It is not proven that the magnitude of a correlation function can be expressed as a sum of exponentials; however,
  the square of the magnitude contributes to the variance correlation function and must have a spectral representation as a sum of exponentials.
  Results of Sec.~\ref{sec:decomposition} demonstrate numerically that the magnitude decays exponentially at large times with a ground-state energy equal to half the ground-state energy of the variance correlation function.
  Eq.~\ref{mRDelta}, which further supposes exponential magnitude excited state contamination, is investigated numerically below, see Fig.~\ref{RThWEM}.}

$e^{R_i(t+1)}$ can similarly be split into an approximately decorrelated product.
Performing this split with regions $[0, t-\Delta t]$ and $[t-\Delta t, t+1]$ gives
\begin{equation}
  \begin{split}
    \avg{ e^{R_i(t+1)} } &= \avg{ e^{R_i(0) + \Delta R_i(t - \Delta t, t - \Delta t)}}  \avg{e^{ \Delta R_i(t + 1, \Delta t + 1) } } 
     \left[ 1   + O\left( e^{-\delta E \Delta t} \right) \right]
     \ \ \ \ .
   \end{split}
  \label{meanDeltaRsplit2}
\end{equation}
The common term in both expressions cancels when constructing the magnitude effective mass, allowing us to define
\begin{equation}
  \begin{split}
    \tilde{M}_R(t, \Delta t) &= \ln\left[ \frac{\avg{e^{\Delta R_i(t, \Delta t)}}}{\avg{e^{\Delta R_i(t + 1, \Delta t + 1 }}} \right]
     = M_R(t) + O\left( e^{- \delta E \Delta t} \right)
     \ \ \ \   .
   \end{split}
  \label{mRDelta}
\end{equation}
Identical steps can be applied to the  phase, leading to
\begin{equation}
   \begin{split}
     \tilde{M}_\theta(t,\Delta t) &= \ln\left[ \frac{\avg{e^{i\Delta \theta_i(t, \Delta t)}}}{\avg{e^{i\Delta \theta_i(t + 1, \Delta t + 1) }}} \right]
     = M_\theta(t) + O\left( e^{- \delta E \Delta t} \right)
    \ \ \ \  .
   \end{split}
   \label{mThDelta}
\end{equation}
The same steps can also be applied to the full correlation function $C_i(t) = e^{R_i(t) + i \theta_i(t)}$.
Noting that
\begin{equation}
   \begin{split}
     e^{\Delta R_i(t,\Delta t) + i \Delta \theta_i(t, \Delta t)} = \frac{C_i(t)}{C_i(t - \Delta t)}
     \ \ \ \   ,
   \end{split}
   \label{decomp}
\end{equation}
the analogous relation for the full effective mass takes the simple form
\begin{equation}
  \begin{split}
     \tilde{M}(t,\Delta t) = \ln\left[ \frac{\avg{C_i(t)/C_i(t - \Delta t)}}{\avg{C_i(t+1)/C_i(t - \Delta t)}} \right]
     = M(t) + O\left( e^{-\delta E \Delta t} \right)
     \ \ \ \  .  
   \end{split}
 \label{mDelta}
\end{equation}
The correlation function ratio effective mass estimator $\tilde{M}(t,\Delta t)$ has different statistical properties than the traditional effective mass $M(t)$ when $\Delta t$ is treated as an independent  $t$.
Note that although $\Delta t$ appears in the numerator and denominator of correlation function ratios superficially similarly to $t_J$ in Eq.~\eqref{eq:emdef}, these two parameters induce quite different statistical behavior.
$C_i(t+1)$ in Eq.~\eqref{mDelta} could be replaced by $C_i(t+t_J)$ (with an appropriate $1/t_J$ overall normalization added).
Taking $t_J>1$ increases the time separation between $C_i(t-\Delta t)$ and $C_i(t+t_J)$ in the correlator ratio in the denominator of Eq.~\eqref{mDelta}, resulting in larger statistical uncertainties in effective mass results, and will not be pursued further here.

  The approximate factorization leading to Eq.~\eqref{mDelta} can be understood from a quantum field theory viewpoint without reference to the magnitude and phase individually.
  Inserting a complete set of states in a correlation function at $t-\Delta t$ allows the correlation function to be expressed as a sum of exponentials $e^{-E_n \Delta t}$ times prefactors representing the amplitude for the system being in the $n$-th state at time $t-\Delta t$.
  These prefactors for each $e^{-E_n \Delta t}$ term are proportional to $e^{-E_n(t-\Delta t)}$, enhancing the amplitude for finding the system in its ground state at large $t-\Delta t$.
  In this way, the contribution to the correlation function from the region $[0,t-\Delta t]$ can be thought of as an effective source for the correlation function in the region $[t-\Delta t, t]$ whose ground-state overlap is dynamically improved compared to the overlap of the original source at time zero.
  The prefactors for each $e^{-E_n \Delta t}$ will depend on the structure of this effective source, but the exponents are fixed by the QCD spectrum.
  The factor of $C_i(t-\Delta t)^{-1}$ in Eq.~\eqref{mDelta} can be considered to be a modification of the effective source in the region $[0,t-\Delta t]$.
  The presence of $C_i(t-\Delta t)^{-1}$ will modify the prefactor of each $e^{-E_n \Delta t}$ term, but it should not affect time evolution of the system in the region $[t-\Delta t, t]$.
  This suggests that an effective mass designed to extract the ground state energy from the sum of $e^{-E_n \Delta t}$ terms, as in Eq.~\eqref{mDelta}, should provide the exact ground state mass at large $\Delta t$ up to corrections arising from excited state contributions to the $e^{-E_n \Delta t}$ sum.
  These corrections should decrease exponentially with increasing $\Delta t$ at a rate set by the energy gap between the ground and first excited state in the system of interest.
  The size of this energy gap will be set by the lowest-lying excitation consistent with the quantum numbers of the system, a derivatively-coupled pion for the case of the nucleon,\footnote{  Multi-hadron correlation functions contain additional low-lying excitations that may introduce larger correlation lengths than $m_\pi^{-1}$ associated for instance with near-threshold bound-states. Such multi-hadron systems are outside the scope of this work.}
  leading to the expectation $\tilde{M}(t, \Delta t) = M(t) + O(e^{-m_\pi \Delta t})$.

  It is not straightforward to construct a representation of $C_i(t-\Delta t)^{-1}$ in terms of local quark and gluon operators that would allow a rigorous proof of these statements, and so numerical LQCD calculations are used to investigate the validity of Eq.~\eqref{mDelta}.
  Exponential reduction of systematic error is numerically demonstrated, but at a faster rate than $m_\pi^{-1}$.
  This suggests that the structure of the effective source plays an important role in determining which $e^{-E_n \Delta t}$ terms are appreciable at the large but finite $\Delta t$ accessible to LQCD calculations in the same way that the structure of the source at time zero determines which excited states make appreciable contributions to the standard effective mass at small $t$.

\begin{figure}[!ht]
  \centering
  \includegraphics[width=.6\columnwidth]{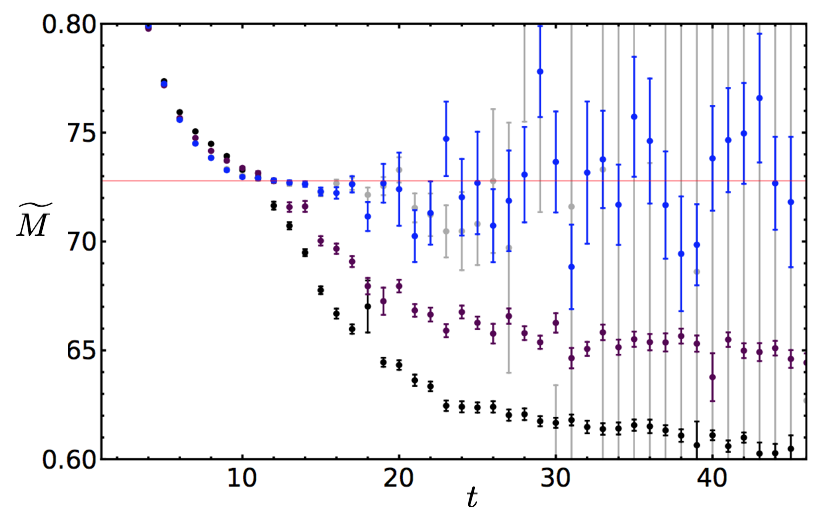}
  \caption{
    Results for the correlation-function-ratio-based estimator $\tilde{M}(t, \Delta t)$. 
  The left panel shows results with $\Delta t =1$ in black, $\Delta t =2$ in purple, and $\Delta t = 8$ in blue, 
  along with the traditional effective mass estimator $M(t)$  shown in gray and a red line at $M_N$ shown for reference. 
  }
  \label{WEM}
\end{figure}
\begin{figure}[!ht]
  \centering
  \includegraphics[width=.6\columnwidth]{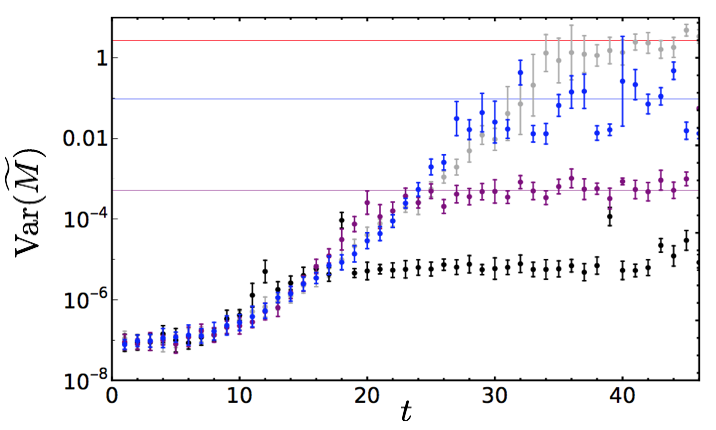}
  \caption{
    Variance in the estimates of $\tilde{M}(t, \Delta t)$ as a function of time $t$ for various choices of $\Delta t$. 
  The black points show $\Delta t = 1$, the purple show $\Delta t = 7$, and the blue show $\Delta t = 15$.
  The gray points show uncertainties in the standard effective mass estimator equivalent to $\Delta t = t$. 
  The purple and blue lines show the expected large-time variance of $\tilde{M}(t, \Delta t)$ with $\Delta t =7,\;15$ 
  predicted by Eq.~\eqref{impscaling} with the overall normalization fixed by the $\Delta t =1$ case. 
  The red line shows the bound of Eq.~\eqref{WEMErrors} with overall normalization again fixed by the $\Delta t =1$ case. 
  Breakdown of statistical inference of broad circular distributions predicts that the large-time variance of $\tilde{M}(t, \Delta t)$ 
  will not systematically rise above the red line for any $\Delta t$.
  }
  \label{WEMErrors}
\end{figure}

The LQCD results for $\tilde{M}_R(t, \Delta t)$ and $\tilde{M}_\theta(t, \Delta t)$ with $\Delta t = 1,\;2\;,8$ are shown in Fig.~\ref{RThWEM}, 
and results for  $\tilde{M}(t,\Delta t)$ are shown in Fig.~\ref{WEM}. 
The statistical uncertainties associated with $\tilde{M}(t, \Delta t)$ are the same as  those of $M(t)$ within the golden window, 
but at large times they become constant in time rather than exponentially increasing. 
This is in accord with our observations about the form of the statistical distributions associated with
$\Delta R_i(t, \Delta t)$ and $\Delta \theta_i(t, \Delta t)$, 
which,  up to small magnitude-phase correlations, indicate that
\begin{equation}
  \begin{split}
    \text{Var}(\tilde{M}(t,\Delta t))  \sim \frac{\text{Var}\left( e^{R_i(t, \Delta t) + i \theta_i(t, \Delta t)} \right)}{\avg{e^{R_i(t, \Delta t) + i \theta_i(t, \Delta t)}}^2} \sim e^{2(M_N - \frac{3}{2}m_\pi)\Delta t}
    \ \ \ \  .
  \end{split}
  \label{impscaling}
\end{equation}
The statistical uncertainties associated with $\tilde{M}(t, \Delta t)$ are constant in $t$, 
although they do increase exponentially with increases in $\Delta t$. 
Since $\Delta \theta_i(t, \Delta t)$ has constant width at large times, 
the inevitable onset of the noise region where statistical inference fails for $\theta_i(t)$ can be avoided. 
The constraint required for reliable statistical inference of $\tilde{M}(t, \Delta t)$ at large times is that the wrapped 
normal variance of $\Delta \theta_i(t, \Delta t)$ can be extracted without large finite sample size errors. 
This constraint can be expressed as a bound on the statistical sample size required for a particular choice of $\Delta t$,
\begin{equation}
  \begin{split}
    N > e^{\sigma^2(\Delta \theta_i(t, \Delta t))} \sim e^{2(M_N - \frac{3}{2}m_\pi)\Delta t}
    \ \ \ \ .
  \end{split}
  \label{WEMstat}
\end{equation}
The statistical uncertainties of $\tilde{M}(t, \Delta t)$ determined from the LQCD correlation functions  are shown in Fig.~\ref{WEMErrors},
from which it can be seen that they become constant at large times for all fixed $\Delta t$. 
For small and moderately large values of $\Delta t = 1,\;7,\;15$, 
the expected exponential increase in large-time statistical uncertainties is observed,
consistent with Eq.~\eqref{WEMstat}. 
Once Eq.~\eqref{WEMstat} is violated, exponential scaling of statistical uncertainties with $\Delta t$ ceases. 
For $\Delta t \lesssim \frac{\ln(N)}{2(M_N - \frac{3}{2}m_\pi)}$, the relative statistical uncertainty in $\tilde{M}(t, \Delta t)$ 
compared to $\tilde{M}(t, \Delta t =1)$ is approximately equal to $N$ rather than $e^{2(M_N - \frac{3}{2}m_\pi)(\Delta t -1)}$.\footnote{These bounds only indicate scaling with $N$. To be made more precise, proportionality constants can be computed using the scaling indicated in Eq.~\eqref{WEMstat}.}
This is seen in Fig.~\ref{WEMErrors} in the large-time behavior of the standard effective mass.

When Eq.~\eqref{WEMstat} is violated, $\Delta \theta_i(t, \Delta t)$ cannot be reliably estimated at large times 
and increasing $\Delta t$ does not improve the accuracy of $\tilde{M}(t, \Delta t)$. 
The standard effective mass estimator can be thought of as evolving with $t \sim \Delta t$, and will become unreliable because of finite sample size effects at large times scaling as $t \gtrsim \ln N / (2(M_N - \frac{3}{2}m_\pi))$.
Similarly, our improved effective mass becomes unreliable for $\Delta t \gtrsim \ln N / (2(M_N - \frac{3}{2}m_\pi))$.
In this extreme case, the bias associated with neglected correlations in $\tilde{M}(t, \Delta t)$ becomes less important than the bias associated with 
statistical inference of overly broad circular random variables. 
Exponential growth of statistical uncertainties with $\Delta t$ suggests that smaller choices of $\Delta t$
where Eq.~\eqref{WEMstat} holds 
likely lead to smaller overall statistical plus systematic uncertainties.

\begin{figure}[!ht]
  \centering
  \includegraphics[width=\columnwidth]{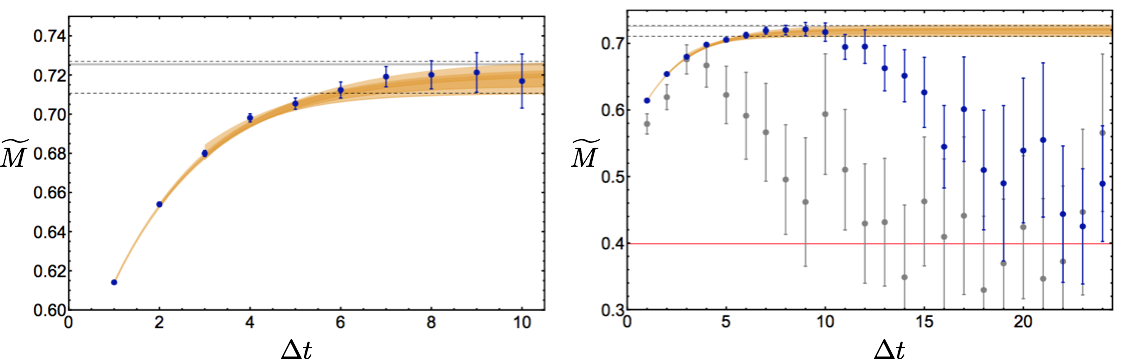}
  \caption{
    In both the left and right panels, results for $\tilde{M}(\Delta t)$ taken from correlated $\chi^2$-minimization fits of $\tilde{M}(t, \Delta t)$ to a constant in the region $25\leq t \leq 40$ with fixed  $\Delta t$ are shown as blue points.
    The tan bands show the results of correlated $\chi^2$-minimization fits of $\tilde{M}(t,\Delta t)$ in various rectangles of $t$ and $\Delta t$ to the three-parameter (constant plus exponential) form shown in Eq.~\eqref{mExtrap}.
  The three light-brown bands all use data from $25 \leq t\leq 40$ and then $1\leq \Delta t \leq 10$, $2\leq \Delta t \leq 10$, and $3\leq \Delta t\leq 10$.
  The black dashed lines show the extrapolated prediction for the nucleon mass including statistical errors from the $2\leq \Delta t \leq 10$ fit added in quadrature with a systematic error calculated as half the maximum difference in central values given by the three fits shown.
 The horizontal gray bands show $M_N \pm \delta M_N$ from the precision NPLQCD calculation of Ref.~\cite{Orginos:2015aya}, which used a high-statistics ensemble of correlation functions with optimized sources generated on the same gauge configurations used here.
 The right panel shows a much larger range of $\Delta t$ and also includes results calculated with a smaller ensemble of $N=5,000$ correlation functions as gray points.
 Deviations from the asymptotic prediction due to finite statistics are clearly visible and lead to incorrect results at much earlier $\Delta t$ in the smaller ensemble.
 }
  \label{WEMSysErrors}
\end{figure}

The systematic bias of $\tilde{M}(t, \Delta t)$ can be explored through calculations at various $\Delta t$. 
Fig.~\ref{WEMSysErrors} shows results for with $\Delta t = 1,\dots,9$. 
For $\Delta t \gtrsim 7$, results for $\tilde{M}(t, \Delta t)$ fit during the large-time noise region $25 \leq t \leq 40$ are statistically 
consistent with fits extracted from the golden window $15 \leq t \leq 25$. 
Late-time fits with $\tilde{M}(t, \Delta t)$ have larger statistical uncertainties than golden window fits. 
More precise fits than either could be made by including both the golden window and the noise region in fits of $\tilde{M}(t, \Delta t)$. 
There is only a minor advantage in including the noisier large-time points in fits that include a precise golden window, 
and this exploratory work does not aim for a more precise extraction of the nucleon mass. 
Practical advantages of large-time fits of $\tilde{M}(t, \Delta t)$ compared to golden window fits of $\tilde{M}(t)$ are more likely to 
be found in systems where a reliable golden window cannot be unambiguously identified. 
Large-time fits of $\tilde{M}(t, \Delta t)$ would also be more advantageous for lattices with larger time directions.

Results for a range of $\Delta t$ shown in Fig.~\ref{WEMSysErrors} can also be used to fit the systematic bias in $\tilde{M}(t, \Delta t)$ and formally extrapolate to the unbiased $\Delta t\rightarrow t\rightarrow \infty$ result. 
During the development of a refined version of this improved estimator~\cite{Wagman:2017xfh},
it was realized that the parametric form of the bias can be deduced by 
considering a decomposition of $[0,t]$ into an extended ``source region'' $[0,t-\Delta t]$ involving $C_i(t)$ and $C_i^{-1}(t-\Delta t)$ and an ``evolution region'' $[t-\Delta t, t]$ only involving $C_i(t)$.
Standard QCD time evolution should apply after the boundary of the source region at $t-\Delta t$, and so at large $\Delta t$ correlation function ratios should scale with $\sim e^{-M_N \Delta t}$ relative to their $t-\Delta t$ boundary values.
Corrections to this ground-state scaling will arise from excited states, which will make contributions to $\left<C_i(t)C_i^{-1}(t-\Delta t)\right>$ scaling as $\sim e^{-(M_N + \delta E)\Delta t}$, where $\delta E$ is the gap between the nucleon ground and first excited state energies.
This allows the dominant contribution to the bias in $\tilde{M}(t,\Delta t)$ to be parametrized as
\begin{equation}
  \begin{split}
    \tilde{M}(t\rightarrow\infty, \Delta t) &= \ln\left[ \frac{e^{-M_N \Delta t}\left( 1 + c\; e^{-\delta E \Delta t} + \dots \right)}{e^{-M_N(\Delta t + 1)}\left( 1 + c\; e^{-\delta E (\Delta t + 1)} + \dots \right)} \right] \\
    & = M_N + c\; \delta E\; e^{-\delta E \Delta t} + \dots,
  \end{split}\label{mExtrap}
\end{equation}
where $c$ is the ratio of excited to ground state overlaps produced by the effective boundary at $t-\Delta t$.
At sufficiently light quark masses and large $\Delta t$, this excited state gap will be set by $m_\pi$.
However, it is noteworthy that Eq.~\eqref{mDelta} involves products of momentum-projected un-averaged correlation functions.
It is familiar from studies of two-baryon correlation functions formed from products of momentum-projected one-baryon blocks that summing over all points in the spatial volume separately for each factor in a product leads to a suppression by $O(m_\pi^{-3}V^{-1})$ in the fraction of points in the product where the nucleons are within one pion Compton wavelength of one another.
It is expected that correlations between $C_i(t)$ and $C_i^{-1}(t-\Delta t)$ described by one-pion excitations will be similarly volume suppressed.
The dominant excited state bias is then expected to arise from excitations that could be produced throughout the lattice volume at the boundary of the source region.
Such excitations are generically far from the nucleon and any other sources of conserved charge, so they should have quantum numbers of the vacuum.
The dominant excited state bias contributing to Eq.~\eqref{mExtrap} is therefore expected to be $e^{-M_\sigma \Delta t}$, where $M_\sigma$ is the mass of the $\sigma$-meson, the lightest excited state with quantum numbers of the vacuum.
Performing a correlated $\chi^2$-minimization three-parameter fit of $\tilde{M}(t, \Delta t)$ to the constant plus exponential form shown in Eq.~\eqref{mExtrap} for noise region data $25 \leq t \leq 40$ gives
\begin{equation}
  \begin{split}
    M_N = 0.7192(49)(42), \hspace{20pt} c = -0.358(26)(17), \hspace{20pt} \delta E = 0.512(65)(73),
  \end{split}\label{mExtrapResults}
\end{equation}
where the first uncertainty is the statistical uncertainty and the second uncertainty is a measure of systematic uncertainty taken from the variation in the central value of the fit as the fitting range in $\Delta t$ is varies. 
The extrapolated result in Eq.~\ref{WEMSysErrors} agrees within uncertainties with the intermediate-time plateau result $M_N = 0.7253(11)(22)$ and with the high-precision GW result $M_N = 0.72546(47)(31)$ of Ref.~\cite{Orginos:2015aya}.
  For the extrapolated large-time result, the total statistical and systematic uncertainties in quadrature is $\delta M_N = 0.0064$, which is larger than the total uncertainty of the plateau region determination $\delta M_N = 0.0025$.
  The large-time plateau considered effectively comprises a two-dimensional region $1 \leq \Delta t \leq 10$ and $25 \leq t \leq 40$ with 150 points.
  The value of the $\chi^2$-minimization fit to this two-dimensional region is most sensitive to points with smaller $\Delta t$ and therefore exponentially smaller uncertainties but is equally sensitive to points with all $t$ that are expected to be approximately decorrelated over intervals $t \gtrsim m_\pi^{-1}$.
  The intermediate-time plateau region $10 \leq t \leq 25$ includes 15 points that are expected to be approximately decorrelated over intervals $t \gtrsim m_\pi^{-1}$.
  The value of the standard effective mass fit is most sensitive to points with smaller $t$ and therefore exponentially smaller uncertainties, though the variance correlation function is not dominated by the three-pion ground state until $t \gtrsim 20$.
  This indicates that results from the intermediate-time plateau have smaller point-by-point uncertainties than points from the large-time noise region.
  The total uncertainty of the noise region result could be reduced by increasing the length of the lattice time direction, while the length of the smaller-time plateau available to standard estimators is restricted by the StN problem.
  The proof-of-principle calculation presented here demonstrates that accurate results can be extracted from the noise region.
  In remains to be seen in future calculations of single- and multi-baryon systems optimized for large-time analysis whether the methods introduced in this work can be used to achieve significantly higher precision with the same resource budget as calculations optimized for smaller-time analysis.

\begin{figure}[!ht]
  \centering
  \includegraphics[width=\columnwidth]{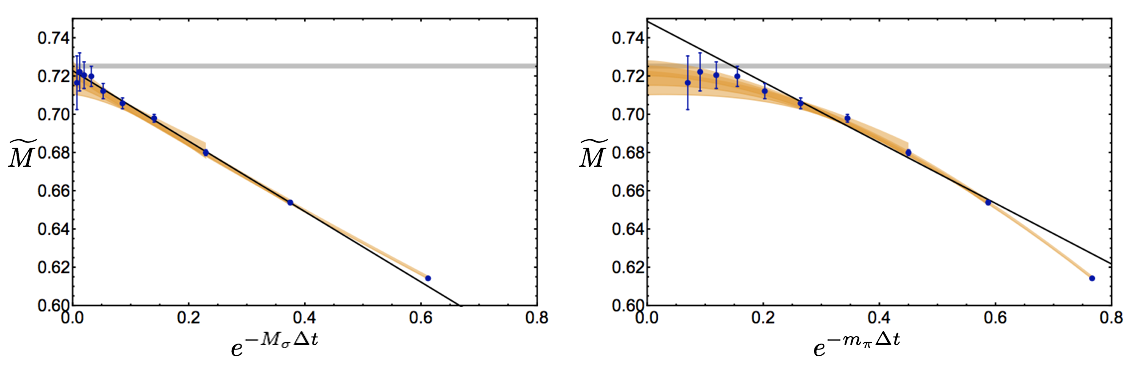}
  \caption{
    The blue points and light-brown bands show the same $\chi^2$-minimization fit results to large-time $\tilde{M}(t,\Delta t)$ plateaus as Fig.~\ref{WEMSysErrors}.
    The horizontal axis has been rescaled to coordinates that would show a linear bias for excited state contributions from $\sigma$-mesons, left, and pions, right.
    Black lines show the central values of $\chi^2$-minimization fits to constrained versions of Eq.~\eqref{mExtrap} where $\delta E$ is fixed to be $M_\sigma$, left, or $m_\pi$, right.
    The horizontal gray bands correspond to $M_N \pm \delta M_N$ from the high-precision NPLQCD calculation of Ref.~\cite{Orginos:2015aya}.
 }
  \label{WEMSysErrors2}
\end{figure}

  The best-fit excitation scale $\delta E = 866(110)(124)$ MeV in Eq.~\ref{WEMSysErrors} can be compared with the $\sigma$-meson mass extracted from mesonic sector calculations to test the heuristic arguments above that lighter excitations will make volume-suppressed contributions.
  Calculations of the $\sigma$-meson face a severe StN problem, 
  particularly at light quark masses where the $\sigma$-meson describes a broad $\pi\pi$ isoscalar resonance rather than a compact QCD bound state,
  but a recent calculation by the Hadron Spectrum collaboration has precisely determined $M_\sigma = 758(4)$ MeV at $m_\pi \sim 391$ MeV where the $\sigma$-meson is weakly bound~~\cite{Briceno:2016mjc}.
  Similarly precise results at slightly higher quark masses are not available for interpolation to $m_\pi \sim 450$ MeV, but a crude extrapolation can be made using the Hadron Spectrum result and the (real part of the) physical position of the $\sigma$-meson pole obtained from dispersive analysis of experimental data: $M_\sigma = 457(14)$ MeV~\cite{Caprini:2005zr,GarciaMartin:2011jx}.
  An extrapolation linear in the pion mass gives $M_\sigma \sim 830$ MeV at $m_\pi \sim 450$ MeV, in rough agreement with the best-fit excitation scale determined above.
  This agreement is insensitive to the form of the extrapolation used, as the Hadron Spectrum $\sigma$-meson mass result at $m_\pi \sim 391$ MeV is itself less that one standard deviation smaller than the best-fit nucleon excitation scale.
  Fits where $\delta E = M_\sigma$ is explicitly assumed can be performed more precisely and lead to consistent results with smaller uncertainties for the nucleon mass $M_N = 0.7226(18)$, as shown in Fig.~\ref{WEMSysErrors2}.
  These fits provide another consistency check on $\delta E$ but do not appropriately capture the systematic uncertainties of explicit assumptions about the excited state spectrum.

The improved estimator proposed here exploits physical locality and finite correlation lengths to extract the effective mass
 from the evolution of $C_i(t)$ between times $t - \Delta t$ and $t$ rather than the full evolution between source time $t=0$ and sink time $t$. 
 The correlation function at time $t - \Delta t$ is effectively treated as a new source so that the effective source/sink separation is fixed 
 to be a constant length $\Delta t$ rather than an increasing separation $t$. 
 The effective source at $t - \Delta t$ still incorporates the dynamical evolution of the system between time 0 and $t - \Delta t$, 
 and in particular has exponentially reduced excited state contamination in the magnitude compared to the original source. 
 In principle $t$ can be taken arbitrarily large with $\Delta t$ fixed in order to extract a plateau in $\tilde{M}(t, \Delta t)$ with 
 arbitrarily small excited state contamination in the magnitude and constant statistical uncertainties across the plateau. 
 The length of the lattice time direction becomes the only factor limiting the length of the plateau in this case.

Similar physical ideas underlie the hierarchical integration approach of Ref.~\cite{Luscher:2001up}. 
In that approach, locality is exploited to decompose correlation functions into products of factors that can be computed on 
subsets of a lattice volume with exponentially reduced StN problems. 
Hierarchical integration has been successfully implemented in studies of gluonic observables~\cite{Meyer:2002cd,DellaMorte:2007zz,DellaMorte:2008jd,DellaMorte:2010yp,Vera:2016xpp} and recently explored for baryon correlation functions in the 
quenched approximation~\cite{Ce:2016idq} and beyond~\cite{Ce:2016ajy}. 
For baryon correlators, the method of Ref.~\cite{Ce:2016idq} implements approximate factorization with systematically 
reducible uncertainties, as in the method proposed here. 
The benefits of the two methods are distinct.
  Hierarchical integration also employs standard statistical estimators for observables defined on sub-volumes to determine correlation functions at large $t$ with exponentially slower StN degradation.
  The new estimators introduced here allow data to be extracted from large-$t$ correlation functions with constant StN, but removing all systematic uncertainties requires an extrapolation to large $\Delta t$ with exponential StN degradation of the same severity as the original correlation function.
  Investigations of the compatibility of and relations between these methods are left to future work.
In addition, this method also has similarities to the generalized pencil-of-functions method introduced to LQCD in 
Ref.~\cite{Aubin:2010jc}, where correlation functions involving shifted source and sink times are combined in a variational basis.
 In the generalized pencil-of-functions approach, shifted source and sink times have primarily been investigated to reduce 
 excited-state contamination rather than StN improvement.

In some sense, $\Delta t$ can be considered a ``factorization'' scale in the time direction.
The LQCD calculations are valid for all energy scales below that defined by the inverse lattice spacing, $\pi/a$.
While well-defined, the MC sampling of the path integral and analysis of baryon correlation functions
 fails to converge in the noise region because of the 
quantum fluctuations encountered along the paths from the source to large times, which include many incoherent hadronic volumes.
The new estimator provides exponentially-improved signal extraction at large times through limiting the number of 
contributing hadronic volumes to those within $\Delta t$, 
but  does not provide a complete description of the IR behavior of QCD, introducing a bias in the extracted mass of the nucleon.
An extrapolation in $\Delta t$, using a form motivated by low-energy pion physics, is used to remove this bias.
While  different, this reminds one of matching LQCD calculations to the p-regime of chiral perturbation theory to 
remove finite-spatial-volume effects.
The idea of performing an extrapolation to overcome a sign problem is not new.
It was introduced thirty years ago to deal with the sign problem in MC calculations of modest size nuclei~\cite{Alhassid:1993yd},
and recently used in lattice effective field theory calculations to continuously evolve between the eigenvalues of nuclear many-body systems
described by a Hamiltonian without a sign problem to one that does have a sign problem~\cite{Lahde:2015ona}.

\section{Summary and Conclusions}
\label{sec:conclusion}

This work presents observations about the nucleon correlation function in LQCD that highlight the role of the complex 
phase in the signal-to-noise problem. 
The magnitude is found to have no StN problem and has the large-time scaling 
$\langle |C_i(t)| \rangle  \sim e^{-\frac{3}{2}m_\pi t}$. 
The nucleon log-magnitude, $R_i(t)$, is approximately described by a normal distribution with linearly increasing mean and almost constant variance. 
The complex phase, which gives the direct importance sampling of $C_i(t)$ a sign problem,  has the large-time scaling of approximately 
$\langle e^{i\theta_i(t)} \rangle \sim e^{-(M_N - \frac{3}{2}m_\pi)t}$.
The StN problem arising from reweighting the complex phase of the nucleon correlation function matches the nucleon StN problem.

We present evidence that nucleon correlation functions are statistically described by a nearly decorrelated product of an approximately log-normal magnitude and wrapped normal phase.
Log-normal times wrapped normal complex correlation functions are consistent with the arguments of Endres, Kaplan, Lee, and Nicholson~\cite{Endres:2011jm}, who suggested stable distributed correlation function logarithms may be a generic feature of quantum field theory and pursued a systematic statistical analysis of unitary fermion correlation functions that provides inspiration for this work.
The wrapped normal phase distribution broadens with time, and at large-times cannot be reliably distinguished from a uniform distribution.
A noise region begins at this point where the sample mean phase becomes biased and systematically 
deviates from the true mean phase. 
In contrast, and importantly, 
$\frac{dR_i}{dt}$ and $\frac{d\theta_i}{dt}$ are described by approximately stable and wrapped stable distributions 
respectively that become constant at large times and can be estimated in the noise region with no StN problem. 

It is remarkable that the Euclidean-time derivate of the logarithm of the correlation function is described by a heavy-tailed 
distribution while the logarithm itself is nearly normally distributed at all times. 
Further studies will be needed to understand the dynamical origin, continuum limit behavior, and universality of 
heavy-tailed Euclidean-time evolution of correlation functions in quantum field theory. 
LQCD calculations at finer lattice spacings are needed to explore the continuum limit of the index of stability 
describing time evolution of the nucleon correlation function. 
Perturbative QCD and model calculations will provide useful insights into the dynamical origin of heavy-tailed time 
evolution of the nucleon correlation function. 
Lattice and continuum studies of other quantum field theories are required to understand the universality of 
heavy-tailed Euclidean-time evolution of correlation functions. Implications for real-time evolution are also left for future investigations.

Building on the observation that $\frac{d\theta_i}{dt}$ has constant width at large times, 
we have proposed a new estimator in Eq.~\eqref{mDelta} for the effective mass of the nucleon correlation function 
that relies on statistically sampling ratios of correlation functions at different times. 
This estimator has a StN ratio that is constant in $t$, the source-sink separation time, and the StN problem instead leads to an exponentially degrading StN ratio in $\Delta t$, the difference between the numerator and denominator sink times.
  The independence of $t$ and $\Delta t$ in this estimator allows similarly precise results to be extracted from all sufficiently large $t$ rather than from a window of intermediate $t$ with standard estimators.
  The new estimator effectively includes $\Delta t$ timesteps of time evolution following $t-\Delta t$ timesteps of dynamical source improvement and it includes a systematic uncertainty that must be eliminated by extrapolating to the limit $ \Delta t \rightarrow t \rightarrow \infty$.
  The systematic uncertainty of the new estimator is expected to decrease as $e^{-\delta E \Delta t}$ for large $\Delta t$, where $\delta E$ is the energy gap between the ground state and the first excited state with appropriate quantum numbers and appreciable overlap with the effective source at $t - \Delta t$.
  Statistical uncertainties increase with increasing $\Delta t$ as $\sim e^{2(M_N - \frac{3}{2}m_\pi)\Delta t}$.
  For $\Delta t \gtrsim \frac{\ln(N)}{2(M_N - \frac{3}{2}m_\pi)}$ additional systematic uncertainties associated with finite-sample-size effects in statistical inference of circular random variables leads to unreliable results in the same way that $t \gtrsim \frac{\ln (N)}{2(M_N - \frac{3}{2}m_\pi)}$ leads to unreliable results in the noise region of standard estimators.

The properties of the new estimator that we have introduced may prove advantageous in the analysis of LQCD 
calculations of nuclei.
Such systems are plagued by a reduced golden window compared to the single nucleon, presently limiting the 
length of plateaus from which to extract energy eigenvalues.
A re-analysis of existing nuclear correlation functions generated by the NPLQCD 
collaboration~\cite{Beane:2012vq,Beane:2013br,Orginos:2015aya} 
is planned in order to determine the utility of this work for such systems.
Binding momenta and other scales appearing in multi-body hadronic systems may affect the form of the extrapolation used to remove the bias of the new estimator.
It will be important to verify and further understand the scaling of the new estimator with pion mass and lattice spacing, as well as to investigate dependences on smearing scales and other scales appearing in LQCD calculations.
Studies of the vacuum channel including glueballs and scalar mesons and analyses of disconnected diagrams provide additional directions for further studies.
Forming ratios of position space, rather than momentum space, correlation functions may be advantageous in future studies.
Other types of LQCD calculations may also benefit from the new estimator, for instance in the isoscalar meson 
sector and those at non-zero baryon chemical potential.

It is not expected that the statistical properties of $\theta_i(t)$ discussed here and, 
in particular the constant large-time width of $\frac{d\theta_i}{dt}$, are unique to single-nucleon correlation functions. 
If analogous statistical properties apply to generic complex correlation functions in 
quantum field theory, then estimators analogous to Eq.~\eqref{mDelta} can be constructed to extract the spectra 
of complex correlation functions and reweighted complex actions without StN problems. 
It remains to be seen if the approaches developed in this work can be fruitfully applied to other systems in 
particle, nuclear, and condensed matter physics that encounter sign and StN problems.

\vspace*{5mm}

{\it Acknowledgments:}
	We would like to thank 
	David Kaplan, Natalie Klco, Silas Beane, Emmanuel Chang, Aleksey Cherman,  Zohreh Davoudi, 
	William Detmold,
        Dorota Grabowska, 
        Kostas Orginos,
        Alessandro Roggero,
        Phiala Shanahan,  and Brian Tiburzi
	for  interesting discussions.
        We thank the members of the NPLQCD collaboration for producing the gauge configurations and baryon correlation functions used in this work, and in particular thank Emmanuel Chang for data management efforts resulting in easily accessible SQLite correlation function databases critical to the exploratory stages of this work.
	This research was supported in part by the National Science Foundation under grant number NSF PHY11-25915 and
	we acknowledge the Kavli Institute for Theoretical Physics for hospitality 
	during much of this work.
	Much of the post-production analysis of the nucleon correlation functions  for this project was carried out on the 
	Hyak High Performance Computing and Data Ecosystem at the University of Washington, 
	supported, in part, by the U.S. National Science Foundation Major Research Instrumentation Award, Grant Number 0922770,
	and by the UW Student Technology Fee (STF). 
	Calculations were performed using computational resources provided
	by  NERSC (supported by U.S. Department of
	Energy grant number DE-AC02-05CH11231),
	and by the USQCD
	collaboration.  This research used resources of the Oak Ridge Leadership 
	Computing Facility at the Oak Ridge National Laboratory, which is supported 
	by the Office of Science of the U.S. Department of Energy under Contract 
	number DE-AC05-00OR22725. 
	The PRACE Research Infrastructure resources  at the 
	Tr\`es Grand Centre de Calcul and Barcelona Supercomputing Center were also used.
	Parts of the calculations used the Chroma software
	suite~\cite{Edwards:2004sx}.  
	MJS was supported  by DOE grant number~DE-FG02-00ER41132, and  in part by the USQCD SciDAC project, 
	the U.S. Department of Energy through grant number DE-SC00-10337.	
	MLW was supported  in part by DOE grant number~DE-FG02-00ER41132.

%


\bibliography{Noise} 

\end{document}